\documentclass[]{JHEP3}

\usepackage{graphicx}
\usepackage{epsfig}
\usepackage{amssymb}
\usepackage{subfigure}
\usepackage{epstopdf}

\newcommand{\myref}[1]{(\ref{#1})}
\newcommand{\smartpap}{p\hskip-7pt\hbox{$^{^{(\!-\!)}}$}}

\title{
Precision electroweak calculation 
of the production of a high transverse-momentum lepton pair
at hadron colliders
}
\author{C.M.~Carloni Calame \\ 
CERN, Physics Department, TH Unit, 1211 Geneva, Switzerland\\
Email: \email{Carlo.Carloni.Calame@pv.infn.it} 
}
\author{G.~Montagna \\
Dipartimento di Fisica Nucleare e Teorica, 
Universit\`a di Pavia and
INFN, Sezione di Pavia, Via A.Bassi 6, I-27100 Pavia, Italy \\
Email: \email{Guido.Montagna@pv.infn.it}
}
\author{ O.~Nicrosini \\
INFN, Sezione di Pavia, Via A.Bassi 6, I-27100 Pavia, Italy \\
Email: \email{ Oreste.Nicrosini@pv.infn.it}
}
\author{ A.~Vicini \\
Dipartimento di Fisica, Universit\`a degli Studi di Milano and
INFN, Sezione di Milano,
Via Celoria 16, I--20133 Milano, Italy \\ 
Email: \email{Alessandro.Vicini@mi.infn.it}
}


\abstract{
We present a detailed study 
of the production of a high transverse-momentum lepton pair at hadron
colliders, 
which includes the exact  ${\cal O}(\alpha)$ electroweak corrections
properly matched with leading logarithmic effects
due to multiple photon emission, as required by the experiments at the 
Fermilab Tevatron
and the CERN LHC. Numerical results for the relevant observables of single $Z$-boson
production at hadron colliders are presented. The impact of the
radiative corrections
is discussed in detail.
The presence in the proton of a photon density is considered
and the effects of the photon-induced partonic subprocesses are analyzed.
The calculation
has been implemented in the new version of the event generator
$\tt HORACE$, which is available for precision simulations of the
neutral and charged current Drell-Yan processes.
}

        \preprint{
         FNT/T 2007/06\\
         IFUM-903/FT }

\begin{document}

\newcommand{\be}{\begin{equation}}
\newcommand{\ee}{\end{equation}}
\newcommand{\nn}{\nonumber}
\newcommand{\bea}{\begin{eqnarray}}
\newcommand{\eea}{\end{eqnarray}}
\newcommand{\bfig}{\begin{figure}}
\newcommand{\efig}{\end{figure}}
\newcommand{\bc}{\begin{center}}
\newcommand{\ec}{\end{center}}
\def\ad{\dot{\alpha}}
\def\ov{\overline}
\def\hlf{\frac{1}{2}}
\def\qrt{\frac{1}{4}}
\def\as{\alpha_s}
\def\at{\alpha_t}
\def\ab{\alpha_b}
\def\sq2{\sqrt{2}}
\newcommand{\smallz}{{\scriptscriptstyle Z}} %
\newcommand{\mz}{m_\smallz}
\newcommand{\smallw}{{\scriptscriptstyle W}}
\newcommand{\mw}{m_\smallw} 
\newcommand{\gw}{\Gamma_{\smallw}} 
\newcommand{\gz}{\Gamma_{\smallz}} 
\newcommand{\sw}{s_{\smallw}} 
\newcommand{\cw}{c_{\smallw}} 
\newcommand{\sdw}{\sin^2\theta_{\smallw}} 
\newcommand{\cdw}{\cos^2\theta_{\smallw}} 
\newcommand{\sqw}{\sin^4\theta_{\smallw}} 
\newcommand{\smallh}{{\scriptscriptstyle H}}
\newcommand{\mh}{m_\smallh}
\newcommand{\mt}{m_t}
\newcommand{\wh}{w_\smallh}
\newcommand{\oa}{${\cal O}(\alpha)~$} 
\newcommand{\oab}{${\cal O}(\alpha)$} 
\def\th{t_\smallh}
\def\zh{z_\smallh}
\newcommand{\Mvariable}[1]{#1}
\newcommand{\Mfunction}[1]{#1}
\newcommand{\Muserfunction}[1]{#1}
%
%


\renewcommand{\thefootnote}{\fnsymbol{footnote}}

%
%
\setcounter{footnote}{0}
\section{Introduction}
At hadron colliders, such as the Fermilab Tevatron and the CERN LHC,
the production of a high transverse-momentum lepton pair,
known as Drell-Yan process \cite{DY}, plays an important role:
it allows, in the neutral current channel, to study the physics of the
$Z$ boson and, in particular, to determine the effective weak mixing angle
from the measurement of the forward-backward asymmetry \cite{AFB};
in the charged current channel,
a high precision determination of two fundamental parameters
of the Standard Model, namely the mass and the decay width of the $W$
boson, can be obtained~\cite{CDF}; 
it provides, both in the neutral and charged current channel, 
stringent constraints on the density functions which
describe the partonic content of the proton \cite{pdf};
it can be used
as a standard reference process 
and, therefore, as a luminosity monitor of the collider
\cite{DPZ,FM}. Furthermore, it
represents
a background to the search for new heavy gauge bosons
\cite{CDFnewgauge}.

The accuracy in the determination of the theoretical cross section has
greatly increased over the years. The calculation of next-to-leading
order (NLO) QCD
corrections~\cite{AEM}
has been one of the first test grounds of perturbative QCD.
Next-to-next-to-leading order (NNLO)
QCD corrections to the total cross section have been computed
in ref.~\cite{HvNM}, but  differential distributions with the same accuracy
have been obtained only recently in ref.~\cite{ADMP}.
The size of the NNLO QCD corrections and the improved stability of the
results against variations of the renormalization/factorization scales
raises the question of the relevance of the \oa electroweak (EW) radiative
corrections, which were computed, in the neutral current channel,
first considering the gauge-invariant subset of QED corrections
\cite{BKS} and then 
the complete set of \oa EW corrections \cite{BBHSW,Zyk}.

A realistic phenomenological study and the data analysis require
the inclusion of the relevant radiative corrections 
and their implementation into Monte Carlo event generators,
in order to simulate all the experimental cuts 
and to allow, for instance, an accurate determination of the 
detector acceptances.
The Drell-Yan processes are included in the standard QCD Parton Shower
generators $\tt HERWIG$~\cite{HERWIG} and $\tt PYTHIA$~\cite{PYTHIA}.
Recently, there has been important progress to improve the QCD
radiation description to NLO, which has been implemented in 
the code $\tt MC@NLO$~\cite{MC@NLO}.
Another important issue is a reliable description of the 
intrinsic transverse momentum of the gauge bosons, which can be
obtained by resumming up to all orders contributions 
due to multiple soft-gluon radiation. 
The generator $\tt RESBOS$~\cite{BY}, used for data analysis at
Tevatron, includes these effects.

If the inclusion of QCD radiation is mandatory for the simulation of
any process at hadron colliders, one should not neglect the impact of
EW corrections on the precision measurement of some
Standard Model (SM) observables, 
like the $Z$ boson mass and decay width, 
important for detector calibration. 
For instance, the generator $\tt ZGRAD$~\cite{BKS,BBHSW} 
includes the exact \oa EW corrections, which have been
shown to induce a shift on the value of $\mz$ extracted from the
Tevatron data of few hundreds of MeV~\cite{CDF}, 
depending on the experimental set up,
mostly due to final-state QED radiation.
Some event generators
can also account  for multiple-photon radiation: in the published
version of $\tt HORACE$~\cite{CMNT,CMNT2, CMNV1} 
QED radiation in leading-log approximation was
simulated by means of a QED Parton Shower \cite{CarloPS};  the
standard tool $\tt PHOTOS$~\cite{PHOTOS} can be adopted to describe QED
radiation in the $Z$ decay.

Since the Drell-Yan events can be used, in principle, to determine the collider
luminosity at a few per cent level, the theoretical cross section
must be known with the same accuracy, requiring also the inclusion
of \oa EW corrections. Furthermore, the \oa EW contributions give
large corrections to the high tails of the invariant mass and lepton
transverse momentum distributions, because of the presence of large EW
Sudakov logarithms~\cite{BBHSW},
thus affecting significantly the SM normalization in the search for
new heavy gauge bosons.

Neutral and charged current Drell-Yan processes can be combined
to obtain an independent way of measuring the $W$ boson mass:
in fact, it has been shown \cite{GK} that the ratios of $W$ and $Z$ observables,
properly defined, is less sensitive to missing higher-order QCD
effects and to the {\it pdf} uncertainties; on the other hand, these
ratios are sensitive to the value of the $W$ mass and can be exploited
for its precise determination.
The impact of the EW corrections in this framework is relevant.
Preliminary studies \cite{carloguido} have shown that these
corrections do not cancel in the ratio and could play an important role in
the $W$ mass determination according to this method.
It is therefore very important to have a tool to perform a realistic
simulation of the Drell-Yan processes, including \oa and multiple
photon corrections, not only in the charged current channel~\cite{CMNV1}, but also in the 
neutral current case.

The aim of this paper is to present a precision EW calculation of the
neutral current Drell-Yan process, 
which includes the exact \oa EW matrix elements properly matched with
leading logarithmic  higher-order QED corrections in the
Parton Shower approach.
The matching of perturbative corrections with Parton Shower, 
which is a topic of great interest in modern QCD simulations \cite{NS},
has already been illustrated in ref.~\cite{CMNV1}
and is realized along the lines already presented in ref. \cite{BCMNP}. 
The use of the {\it pdf} set $\tt MRST2004QED$~\cite{MRST2004QED}, 
which describes the
partonic content of the proton also in terms of a photon density,
implies the calculation of photon-induced partonic processes. The
latter contribute to the inclusive cross section
$\sigma\left(p\smartpap \to l^+l^- (n\gamma)+X\right)$
and are of the same perturbative order as the Born
Drell-Yan tree-level reaction (subprocess $\gamma\gamma\to l^+l^-$)
or of the same perturbative order of the \oa corrections
(subprocess $\gamma q\to l^+l^- q$).
The latter have been recently addressed to in ref.~\cite{Arbuzov:2007kp}.

Several distributions of physical interest are analyzed,
disentangling the effect of different classes 
of radiative corrections and discussing various sources of theoretical
uncertainty.
The calculation is implemented in the new version of the 
Monte Carlo event generator $\tt HORACE$~\footnote{The code can be downloaded from the url
{\tt http://www.pv.infn.it/hepcomplex/horace.html}},
which combines, in a unique tool,
the good features of the QED Parton Shower approach with the 
additional effects
present in the exact \oa EW calculation.
This task is non trivial from several technical points of view and
faces all the conceptual problems of developing a NLO event generator.

The paper is organized as follows.
In Section~\ref{oa} we present the calculation 
of the partonic subprocesses and of \oa radiative corrections.
In Section~\ref{matching} we describe the matching of the fixed-order
results with the QED Parton Shower.
In Section~\ref{hadronlevel} we present the computation of the hadron-level
cross section $\sigma\left(p\smartpap \to l^+l^- (n\gamma)+X\right)$
and discuss the subtraction of the initial-state collinear
singularities to all orders.
In Section~\ref{results}
we present phenomenological results for several physical distributions
and discuss the impact of \oa EW  corrections, photon-induced processes and higher-order QED
contributions. Finally, in Section~\ref{concllabel} we draw our
conclusions and discuss possible developments of this work.

\section{Partonic process: matrix elements calculation}
\label{oa}
We consider a description of the proton 
which includes also the presence of a photon as a parton, 
described by the corresponding density function, 
as done for instance in ref.~\cite{MRST2004QED}.
The lowest-order cross section for the production of a high transverse-momentum 
lepton pair starts at ${\cal O}(\alpha^2)$
and receives contributions from two partonic subprocesses, namely
$q(p_1)~ {\bar q(p_2)} \to l^-(p_3)~ l^+(p_4)$
and also
$\gamma(p_1)~ \gamma(p_2) \to l^-(p_3)~ l^+(p_4)$.
When considering this final state at ${\cal O}(\alpha^3)$,
we must include, in addition to the usual radiative process
$q(p_1)~ {\bar q(p_2)} \to l^-(p_3)~ l^+(p_4) \gamma(p_5)$, 
also a new subprocess, i.e.
$q(p_1)~ \gamma(p_2) \to l^-(p_3)~ l^+(p_4) q(p_5)$,
which contributes to the inclusive signature.
\begin{figure}
\begin{center}
\includegraphics[scale=0.8]{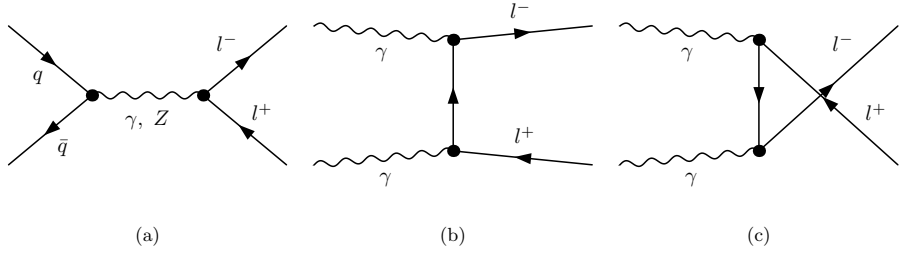}
\caption{Born diagrams for the $q{\bar q}$ (a) and for the
  $\gamma\gamma$ (b,c) subprocesses.}
\label{figborn}
\end{center}
\end{figure}

\subsection{Born approximation}
We start considering the neutral current Drell-Yan partonic process
$q(p_1)~ {\bar q(p_2)} \to l^-(p_3)~ l^+(p_4)$, which is depicted in
figure \ref{figborn} (a).
This process is a neutral current process
and its amplitude, neglecting the Higgs-boson contribution,
is mediated by $s$-channel photon and $Z$-boson exchange.
In the unitary gauge, the tree-level amplitude reads as
\bea
{\cal M}_0 &=& {\cal M}_\gamma + {\cal M}_Z 
\label{M0}\\
{\cal M}_\gamma &=& 
-~e^2~Q_q Q_l
~\frac{g_{\mu\nu} - k_\mu k_\nu/s}{s} 
\left[
{\bar v}(p_2) \gamma^{\mu} u(p_1)
\right]
\left[
{\bar u}(p_3) \gamma^{\nu} v(p_4)
\right]\nonumber \\
&\equiv &
-~e^2~Q_q Q_l
~\frac{g_{\mu\nu} - k_\mu k_\nu/s}{s} 
J_{em}^{\mu} J_{em}^{\nu}\nonumber\\
{\cal M}_Z &=& 
-~\frac{e^2}{s_\theta^2 c_\theta^2}
~\frac{g_{\mu\nu} - k_\mu k_\nu/s}{s-\mz^2+i\gz\mz} 
\left[ {\bar v}(p_2) 
\left( v_q~\gamma^{\mu}+a_q \gamma^{\mu} \gamma_5 \right) 
        u(p_1) \right] 
\left[ {\bar u}(p_3) 
\left( v_l~\gamma^{\nu}+a_l \gamma^{\nu} \gamma_5 \right) 
       v(p_4) \right]\nonumber\\
&\equiv &
-~\frac{e^2}{s_\theta^2 c_\theta^2}
~\frac{g_{\mu\nu} - k_\mu k_\nu/s}{s-\mz^2+i\gz\mz} 
J_{Z,q{\bar q}}^{\mu} J_{Z,l^+l^-}^{\nu}\nonumber
\eea
where 
$\mz$ is the $Z$-boson mass and $\gz$ is the $Z$ decay width,
necessary to describe the $Z$ resonance region,
$s=(p_1+p_2)^2$ is the squared partonic center-of-mass (c.m.) energy
and
$k^\mu=p_1^\mu+p_2^\mu$,
$\alpha=e^2/(4\pi)$ is the fine structure constant,
$c_\theta \equiv \mw/\mz$ is the cosine of the weak mixing angle. 
The vector and axial-vector couplings of the $Z$-boson to fermions are
$v_f=T_f-2 Q_f s_\theta^2$ and $a_f=-T_f$ 
where $T_f=\pm 1/2$ is the third component of the weak
isospin and $Q_f$ is the electric charge of the fermion $f$.

The subprocess 
$\gamma(p_1)~ \gamma(p_2) \to l^-(p_3)~ l^+(p_4)$,
which is depicted in
figure \ref{figborn} (b,c),
is, at lowest order, a pure QED reaction, whose differential cross section, 
in the partonic c.m. frame and neglecting all fermion masses, reads as
\bea
\frac{d\hat\sigma_{\gamma\gamma}}{d\cos\theta}&=&
\frac{2\pi\alpha^2}{s}
\left(
\frac{1 + \cos^2\theta}{\sin^2\theta}
\right)
\eea

\subsection{The \oa calculation}
The complete  \oa EW corrections to the neutral current Drell-Yan process
have already been computed in refs.~\cite{BBHSW,Zyk}.
We have repeated independently the calculation
and included in addition the photon-induced processes.
We summarize here the main features of our approach.

The \oa corrections include the contribution of real and virtual corrections.
The virtual corrections follow from the perturbative expansion
of the $2\to 2$ scattering amplitude
${\cal M} = {\cal M}_0 + {\cal M}_{\alpha}^{virt} + \cdots$ and contribute,
at \oab, with $2 {\mathrm Re}({\cal M}_{\alpha}^{virt} {\cal M}_0^*)$.  
The \oa virtual amplitude includes two contributions, namely the 
one-loop renormalization of the tree-level amplitude and  the
virtual one-loop diagrams. The real corrections are due to the
emission of one extra real photon and represent the lowest order of
the radiative process
$q(p_1) {\bar q(p_2)} \to l^-(p_3) l^+(p_4) \gamma(k)$.
They can be further divided into soft and hard corrections,
${\cal M}_{1}={\cal M}_{1}^{soft}+{\cal M}_{1}^{hard}$.
The former satisfies, by definition, the Born-like $2\to 2$ kinematics
and can be factorized as 
$|{\cal M}_{1}^{soft}|^2 = \delta_{SB} |{\cal M}_0|^2$, 
where $\delta_{SB}$ is a universal factor that depends only on the
properties of the external particles.
The total cross section includes soft and hard corrections
and is independent of the
cut-off used to define the two energy regions.
Virtual and real soft corrections are separately divergent due to the
emission of soft photons, but the divergence cancels in the sum of the
two contributions.

\subsubsection{Virtual corrections}
\begin{figure}
\begin{center}
\includegraphics[scale=0.8]{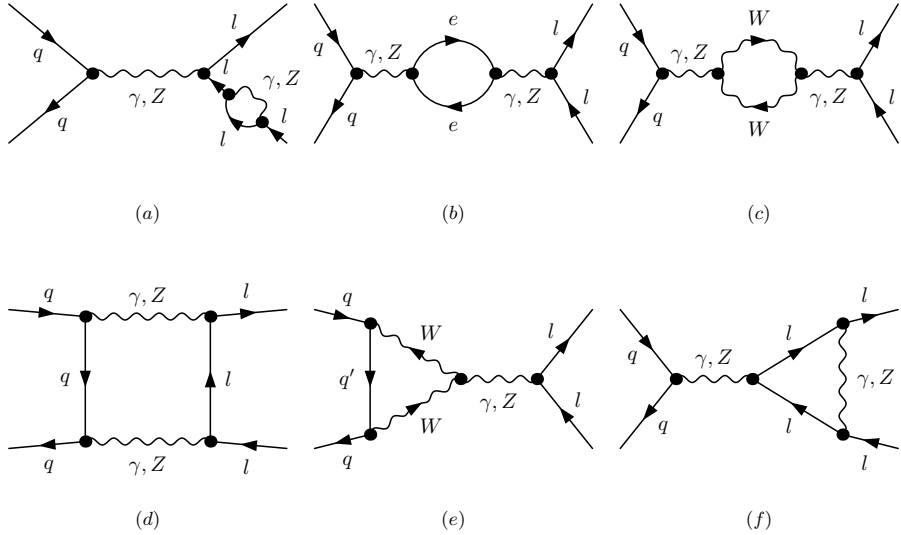}
\caption{Some examples of one-loop virtual diagrams.}
\label{figvirt}
\end{center}
\end{figure}

The \oa virtual corrections to a $2\to 2$ reaction include the
contribution of counterterms, self-energy, vertex and box corrections.
Few diagrams representative of the different kinds 
of corrections are depicted in figure~\ref{figvirt}.
The \oa virtual corrections
have been calculated using the packages $\tt FeynArts$ and
$\tt FormCalc$~\cite{FA,LT}.
The numerical evaluation of the 1-loop integrals has been done using
the package $\tt LoopTools2$~\cite{LT}, based on the library
$\tt ff$~\cite{ff}.
We will write the 1-loop virtual amplitude as
${\cal M}_{\alpha}^{virt}=
{\cal M}_{\alpha}^{cts}+
{\cal M}_{\alpha}^{self}+
{\cal M}_{\alpha}^{vertex}+
{\cal M}_{\alpha}^{box}$,
where ${\cal M}_{\alpha}^{cts}$ includes all the counterterms and the
wave function corrections on the external legs,
${\cal M}_{\alpha}^{self}$ describes the self-energy corrections to
the photon and to the $Z$ propagator and the contribution due to the
$\gamma-Z$ mixing and
${\cal M}_{\alpha}^{vertex,box}$ describe respectively the vertex and
the box corrections. 
The mass of the fermions in the scalar 1-loop
integrals regularizes in a natural way the mass singularities due to the
emission of a (virtual) collinear photon.
The infrared divergence of the integrals has been regularized by means
of a small photon mass $\lambda$.

The introduction of the $Z$ decay width in the propagator of the $Z$
boson is mandatory to describe the resonance region and to regularize the
divergence due to the pole of the propagator. 
In the 't Hooft-Feynman gauge, 
the 1-loop corrections to the $Z$ propagator
read
\be
(-i g^{\mu\nu})\frac{1}{s-\mz^2+i\gz\mz}
\left({\mathrm Re}(\Pi_{ZZ}(s))+\delta\mz^2+(s-\mz^2)\delta Z_\smallz\right)
\frac{1}{s-\mz^2}
\label{Zprop}
\ee
where $\Pi_{ZZ}(s)$ is the transverse part of the $Z$ self-energy corrections,
$\delta\mz^2$ and $\delta Z_\smallz$ are  the $Z$ mass
and wave function renormalization constants, respectively.
The two counterterms cancel the divergences present in the self-energy
corrections. 
We remark that we have to take the real part of $\Pi_{ZZ}(s)$
in order to avoid a double counting with the decay width in the
denominator.
In fact the \oa contribution is 
$2 {\mathrm Re}({\cal M}_{ZZ}^{self} {\cal M}_0^*)$:
the real part prescription in the latter expression avoids the double
counting in the interference with the $Z$ tree-level diagram,
but not the one in the interference with the photon tree-level diagram.
Furthermore, the second factor $1/(s-\mz^2)$ in
eq.~\myref{Zprop} is not corrected by the decay width, 
again to avoid a double counting; 
we can check, by expanding the self-energy corrections around
$s=\mz^2$, that the \oa expression is regular for $s\to\mz^2$.

The $\gamma-Z$ mixing diagrams 
require the introduction in the $Z$ propagator
of a decay width to describe the $Z$-resonance region, i.e.
\be
(-i g^{\mu\nu})\frac{1}{s-\mz^2+i\gz\mz}
\left(
{\mathrm Re}(\Pi_{\gamma Z}(s))
+\frac{(s-\mz^2)}{2} \delta Z_{\gamma\smallz}
+\frac12 \delta Z_{\smallz\gamma}
\right)
\frac{1}{s}
\label{gammaZprop}
\ee
Also in this case we need to take the real part of $\Pi_{\gamma Z} $
to avoid the double counting between the imaginary part of the
self-energy and the imaginary term proportional to the decay width.

Examples of vertex and box corrections are depicted in figure~\ref{figvirt}.
The abelian vertex diagrams and the box diagrams with a photonic correction
are infrared divergent.
All the vertex and box diagrams with a $Z$-boson exchange
yield the so-called EW Sudakov logarithms, namely terms like 
$\alpha\log^2\left(s/\mz^2 \right)$, whose importance grows for large
invariant mass of the final-state lepton pair, while they are almost
negligible at the $Z$ resonance.
The box diagrams with a photon and a $Z$ boson in the loop yield a
logarithmic divergence at $s=\mz^2$.
Since all the terms with $\log(s-\mz^2)$ form a gauge invariant 
subset of corrections,
we regularized the divergence by replacing everywhere
$\log(s-\mz^2)\to\log(s-\mz^2+i\gz\mz)$.
The $\gamma Z$ boxes are resonant at $s\sim\mz^2$. The divergence has
been regularized by modifying the box amplitude in the following way
\bea
{\cal M}^{box} &=&
{\cal M}^{box}_{\gamma\gamma}  +
{\cal M}^{box}_{\gamma Z}  +
{\cal M}^{box}_{Z Z}   \to \nonumber \\
\to &&
{\cal M}^{box}_{\gamma\gamma}  +
\frac{s-\mz^2}{s-\mz^2+i\gz\mz}{\cal M}^{box}_{\gamma Z}  +
{\cal M}^{box}_{Z Z} \label{boxrescale}
\eea

The calculation has been repeated with two different gauge choices,
namely the $R_{\xi}$ with $\xi=1$ gauge and the background field
gauge, with parameter $q=1$~\cite{BFG}.
The two results perfectly agree, and this is an important check of the
calculation of
the bosonic self-energy and of the non-abelian vertex corrections.

Concerning the renormalization of the 1-loop amplitude,
the ultraviolet divergences which appear from the
virtual diagrams can be cancelled with the mass, $\delta\mz^2$, and wave
function, $\delta Z_{\smallz}$, renormalization constants of the $Z$
boson,
with the wave function renormalization of the photon, 
$\delta Z_{\gamma\gamma}$, and of the $\gamma Z$ mixing,
$\delta Z_{\gamma \smallz}$ and $\delta Z_{\smallz\gamma}$
and 
by the renormalization of the two vertices $Zq{\bar q}$ and $Zl^+l^-$. 
The latter include the charge renormalization and the wave
function renormalization of the external fermions, of the photon  and
of the $Z$ boson. 
We use the electric charge and the gauge boson masses $e,\mw,\mz$ 
as input parameters, 
we replace the bare coupling $e_0$ in terms of the renormalized quantity and of
the corresponding counterterm $e_0 = e (1-\delta e/e)$.
The electric charge counterterm is fixed by the request that in
Thomson scattering the renormalized charge is given by the fine
structure constant;
its expression depends on the quark masses running in the photon
vacuum polarization, the
value of which can be adjusted in order to make the running
electric charge reproduces the value $\alpha(\mz^2)$~\cite{fred}.
For the sake of brevity, we can introduce a counterterm for the weak
mixing angle, which is given by  a combination of
the mass counterterms of the $W$ and $Z$ bosons, following from its
definition.
The $W$- and $Z$-boson mass and wave function renormalization constants,
the $\gamma\gamma$ and $\gamma Z$ wave function renormalization
constants are defined in the on-shell scheme.

The choice of the input parameters of the SM lagrangian has a numerical impact
on the calculation of the physical observables. 
In order to obtain predictions with the smallest possible parametric
uncertainty, it would be convenient to choose as input parameters
$\alpha,G_{\mu},\mz$, which was the common choice at LEP1.
The drawback is that in this approach $\mw$ is a prediction of the SM,
whereas, at hadron colliders, 
it would be interesting to keep it as an input, 
which can be measured fitting the data.
It is therefore more useful to choose $\alpha, \mw, \mz$ as inputs.
For instance, 
the method proposed in ref.~\cite{GK} to study the ratio of $W$ and $Z$
observables
and to use it to determine the $W$ mass
requires $\mw$ as input of both charged and neutral current
simulations, which will then be compared with the data.

Unfortunately, the choice $\alpha, \mw, \mz$ suffers, if
strictly applied, of the ambiguity due to the choice of the quark mass
values in the photon vacuum polarization.
The ambiguity can be removed
by re-expressing, in the Born amplitude, the fine structure constant in
terms of a different quantity, like the Fermi constant $G_\mu$ or
the effective electromagnetic coupling at the scale $q^2$,
$\alpha(q^2)$, whose \oa expressions exactly cancel the quark mass
dependence due to the vacuum polarization diagrams.
In order to rewrite the tree-level amplitude in terms of these new couplings,
we follow the prescriptions developed at LEP1 to study the $Z$-resonance
and treat separately the photon- and the $Z$-diagram contributions.
In ${\cal M}_{\gamma}$ we replace 
\bea
e^2&\to& e^2(q^2) = e^2/\left(1-\Delta\alpha(q^2)\right)\label{dalpha}\\
\Delta\alpha(q^2)&=&\Pi^{(f)}_{\gamma\gamma}(q^2)-\Pi^{(f)}_{\gamma\gamma}(0)
\nonumber
\eea
where $\Delta\alpha(q^2)$ expresses the running of the electric charge due
to the fermionic photon vacuum polarization 
($\Pi^{(f)}_{\gamma\gamma}(q^2)$
is related to the transverse part of the photon
self-energy by 
$A^{(f)}_{\gamma\gamma}(q^2)=q^2 \Pi_{\gamma\gamma}^{(f)}(q^2)$ ).
The quantity $\Delta\alpha$ includes also the non-perturbative hadronic contribution
due to a loop of quarks at low virtualities. 
For the evaluation of
$\Delta\alpha^{(5)}_{\rm had}$ we use the routine  {\tt hadr5n} \cite{fred}.
Because we include the photon vacuum polarization effects in the lowest-order 
coupling, we have to  subtract the \oa expansion of
$e^2(q^2)$, to avoid a double counting when we include the full set
of \oa corrections.

In the case of ${\cal M}_Z$ we can rewrite 
$e^2/(s_\theta^2 c_\theta^2)$ as $g^2/c_\theta^2$ and
then use the relation, computed up to \oab, 
of the weak coupling $g$ with the Fermi constant and the $W$-boson mass
\be
\frac{G_{\mu}}{\sqrt{2}}=\frac{g^2}{8\mw^2}\left(1+\Delta r\right)
\label{gmu}
\ee
The quantity $\Delta r$ represents all the radiative corrections to the
muon-decay amplitude~\cite{Sirlin80}.

\subsubsection{Bremsstrahlung corrections}
\begin{figure}
\begin{center}
\includegraphics[scale=0.95]{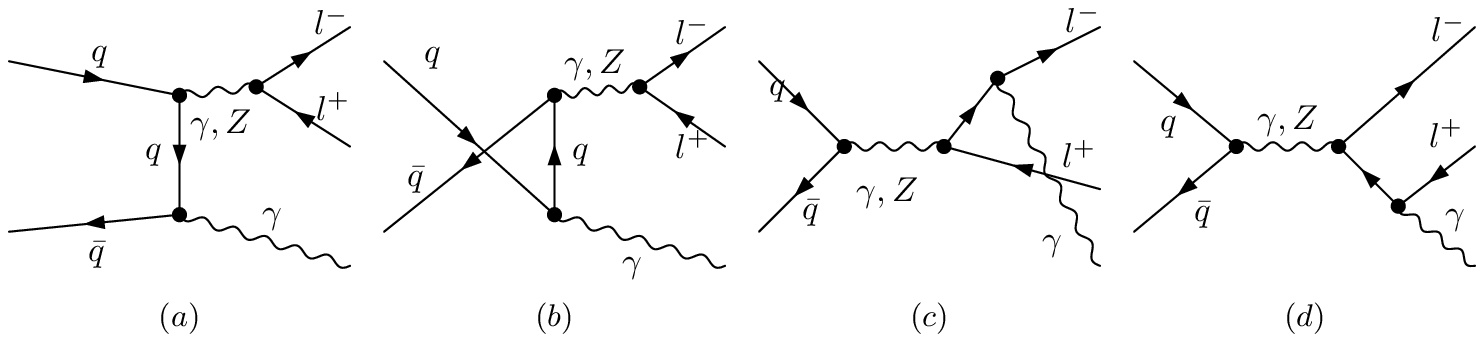}
\caption{\oa bremsstrahlung diagrams.}
\label{figreal}
\end{center}
\end{figure}

The real radiative corrections to the neutral current Drell-Yan process,
described by the amplitude ${\cal M}_{1}$,
are given by all the Feynman diagrams (figure~\ref{figreal})
with the emission of one extra
photon from all the electrically charged legs of the Born diagrams.

The probability amplitude has been calculated in the unitary gauge
with massive fermions.
We integrate the squared matrix element 
over the whole photon phase space and split the allowed
photon energy range in two intervals,
$[\lambda,\Delta E]$ and $[\Delta E,  E_{max}]$. 
The cut-off $\Delta E\ll \sqrt{s}$ is chosen in such a way that the photon
with smaller energy is considered soft and does
not modify the $2\to 2$ kinematics of the Born amplitude.
The small photon mass $\lambda$ has been introduced to regularize
the infrared divergence.
In this energy region the phase space integral,
including the full angular integration, can be solved analytically.
The result can be expressed in a factorized form, as
\be
\int_{\Omega} \frac{d^3{\mathbf k}_{\gamma}}{(2\pi)^3 2 E_{\gamma} }
|{\cal M}_{1}|^2 
~=~
|{\cal M}_0|^2~~\sum_{f=q,{\bar q},l^+l^-}~\delta_{SB}(f,\lambda)
\ee
where the soft Bremsstrahlung factor, see e.g. ref.~\cite{Denner},
depends on the mass and electric charge of the external
radiating particles
and the phase-space region $\Omega$ is defined by the request that
the photon energy $E_{\gamma}$ satisfies
$\lambda\leq E_{\gamma} \leq \Delta E$.
We have explicitly checked that the sum of the virtual and soft-real
contributions is independent of the choice of the photon mass
$\lambda$, in the limit of small $\lambda$ values.

In the hard energy region the phase-space integration has been
performed numerically, with Monte Carlo techniques
improved by importance sampling to take care of collinear and infrared
singularities, as well as the peaking behaviour around the $Z$ resonance.
The sum of the soft and of the hard photon cross sections
is independent of the cut-off $\Delta E$.
We have checked the independence of our numerical results from the
choice of the infrared separator $\varepsilon\equiv \Delta E/E$
for $10^{-8} \leq \varepsilon \leq 10^{-4}$.


\subsubsection{Photon-induced processes}
\begin{figure}
\begin{center}
\includegraphics[scale=0.95]{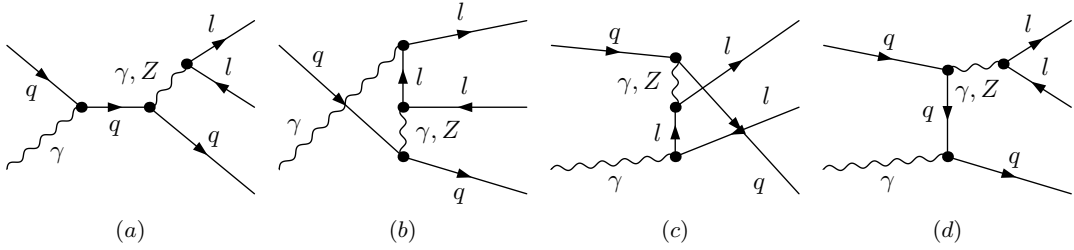}
\caption{Photon-induced process diagrams.}
\label{figinduced}
\end{center}
\end{figure}

In ref.~\cite{MRST2004QED} it has been proposed 
a new parametrization of the partonic content of the proton,
which also includes a photon probability density.
When using this set of {\it pdf},
the inclusive cross section
$\sigma\left(p\smartpap \to l^+l^-+X \right)$
receives contributions also from the partonic subprocesses
$q(p_1) \gamma(p_2) \to l^+(p_3) l^-(p_4) q(k) $ (photon-induced),
depicted in figure \ref{figinduced}.
The latter are of the same perturbative order as the real
bremsstrahlung corrections described in the previous subsection,
i.e. they are an \oa correction to the Born process of eq.~\myref{M0}.
The squared amplitude of the photon-induced processes can be obtained 
by crossing symmetry from the real bremsstrahlung one,
evaluating the latter with the exchange $(p_2 \leftrightarrow -k)$
and multiplying the result by a $(-1)$ factor to account for the exchange of a
fermionic line.



\subsection{Higher-order electroweak effects}
\label{IBA}
To incorporate higher-order EW corrections in a Born-like expression written with effective
couplings,
we followed the approach of ref.~\cite{DS91}, where the tree-level amplitude
has been improved and takes into account all the self-energy and vertex corrections.
The latter have been included by defining 
an effective overall coupling and an effective weak mixing angle.

The amplitude ${\cal M}_Z$ becomes
\be
{\cal M}_Z = \frac{i 8 \, G_{\mu} \mz^2 }{\sqrt{2}}~
\frac{\rho_{fi}(q^2) }{1-\delta\rho_{irr}}~
\frac{J_{Z,q{\bar q}}\cdot J_{Z,l^+l^-}  }{ q^2-\mz^2+i\gz\mz}
\label{eqIBA}
\ee
where the coupling $v_f$ of eq.~(\ref{M0}) is replaced by
$\tilde v_f = T_f-2 Q_f \kappa_f(q^2) s_\theta^2$.
The definition of the quantities $\rho_{fi}, \delta\rho_{irr}, \kappa_f(q^2)$
can be found in ref.~\cite{DS91}.
Eq.~(\ref{eqIBA}) incorporates also higher-order effects beyond \oab,
because of the resummation of $\delta\rho_{irr}$
and of the fermionic part of the $Z$ self-energy contained in
$\rho_{fi}$.
Furthermore, $\delta\rho_{irr}=\delta\rho_{irr}^{(1)}+
\delta\rho_{irr}^{(2)}$ contains also leading two-loop corrections.
In the amplitude ${\cal M}_\gamma$ we replace the fine structure
constant with the running electromagnetic coupling according to eq.~(\ref{dalpha}). 

For the numerical analysis we combine the exact \oa corrections described
in the previous subsections and evaluated with $\alpha(0),\mw,\mz$ as
input parameters, with the Born-like expressions of this section.
In order to avoid a double counting in the combination,
we subtract the \oa content of eq.~(\ref{eqIBA}). All the numerical results shown 
in the following are obtained according to the EW scheme here described.

\section{Matching QED higher orders and hadron-level cross section}
\label{matchhadr}
\subsection{Matching}
\label{matching}
In this section we describe the matching of the
\oa EW calculation with
higher-order QED corrections (cfr. ref. \cite{BCMNP}).
The latter can be included in a generic scattering cross section
in the QED Parton Shower approach, which resums to all orders the
leading logarithmic effects.
At \oa the Parton Shower reproduces only the QED leading-log
approximation of the exact \oa EW calculation presented in section
\ref{oa}.
We would like to combine the exact \oa results 
and QED higher orders,
to improve the approximation intrinsic to the Parton Shower,
avoiding at the same time double counting at \oab.
A detailed description of the matching procedure can be found in
refs.~\cite{CMNV1,BCMNP}. Here we recall the basic ideas and results.

Our master formula is
\be
d\sigma^{\infty}=
F_{SV}~\Pi(Q^2,\varepsilon)~
\sum_{n=0}^\infty \frac{1}{n!}~
\left( \prod_{i=0}^n F_{H,i}\right)~
|{\cal M}_{n,LL}|^2~
d\Phi_n
\label{matchedinfty}
\ee
and can be understood in the following way:
1) the tree-level cross-section ($n=0$ in the sum)
is corrected, in leading-log approximation, to all orders
by the Sudakov form factor $\Pi(Q^2,\varepsilon)$, which accounts for virtual
and soft-photon emission up to a scale $\varepsilon$ in a hard-scattering process
characterized by a virtuality scale $Q$; 
2) the resulting expression is dressed by the QED Parton Shower,
with the real bremsstrahlung $n$-photons squared matrix elements in
leading-log approximation;
3) the correction factors $F_{SV}$ and $F_{H,i}$ provide the remaining
\oa corrections missing in the leading-log approximation: in particular, $F_{SV}$
contains the remainder of the soft plus virtual corrections, whereas
$F_{H,i}$ gives, for the real photon emission, 
the correction due to the exact bremsstrahlung matrix element with
respect to the Parton Shower approximation.

The expansion at \oa of eq.~\myref{matchedinfty} coincides
with the exact NLO cross section.
Furthermore, all higher-order leading-log contributions
are the same as in a pure QED Parton Shower.
It is worth noticing that $F_{SV},F_{H,i}$ are, by construction,
infrared safe and free of collinear logarithms.

\subsection{Hadron-level cross section}
\label{hadronlevel}
In this section we discuss the calculation of the hadron-level cross section 
$\sigma(p\smartpap\to l^+l^- +n\gamma +X) $
and the procedure to subtract the initial-state mass
singularities. 
This procedure makes the resulting hadron-level cross section independent of
the (fictitious) value of the initial-state quark masses.

The subtraction at \oa has been discussed, for example, in ref.~\cite{DK} and
is obtained by a redefinition of the parton densities.
The hadron-level cross section at \oa can be written as
\bea
\label{sigmahad}
d\sigma(p\smartpap\to l^+l^-+X)&=&
\sum_{a,b=q,\bar q, \gamma}\int_0^1 dx_1 dx_2~~
q_a(x_1,M^2) q_b(x_2,M^2)
\left[
d{\sigma}_0^{ab}~+~
d{\sigma}_{\alpha}^{ab}
\right]\\
&&-~\left( 
\Delta q_a(x_1,M^2) q_b(x_2,M^2)+
q_a(x_1,M^2) \Delta q_b(x_2,M^2)
\right) \, d\sigma_0^{ab}
\nonumber
\eea
where $a,b$ run over all parton species described by the densities 
$q_i(x,M^2)$, $M$ is the factorization scale,
$d\sigma_0^{ab}$ and $d\sigma_\alpha^{ab}$ are the Born and \oa partonic
cross sections, initiated by partons $a$ and $b$, respectively.
In the case of photon-induced processes, 
$d\sigma_0^{q\gamma}=d\sigma_0^{\gamma q}=0$
\footnote{In the numerical evaluation, we omit \oa corrections 
to the subprocess $d\sigma_0^{\gamma\gamma}$, because numerically negligible.}.

The relevant \oa subtraction terms are
\bea
\Delta q_i(x,M^2)&=&
\int_x^1 dz~
q_i\left(\frac{x}{z},M^2\right)
\frac{\alpha}{2\pi} Q_i^2
\left[
P_{q\to q\gamma}(z)
\left(
\log\left(\frac{M^2}{m_i^2}\right)-2 \log(1-z)-1 
\right)
\right]_+ 
\nonumber\\
&&~~~~
+ f_q(z)+~q_\gamma\left(\frac{x}{z},M^2\right)
\frac{\alpha}{2\pi} Q_i^2
\left[
P_{\gamma\to q\bar q}(z)
\left(
\log\left(\frac{M^2}{m_q^2}\right)
\right)
\right]~+f_\gamma(z)
\label{deltaq}
\\
\Delta q_\gamma(x,M^2)&=&
\sum_{i=q,\bar q}~\int_x^1 dz~
q_i\left(\frac{x}{z},M^2\right)
\frac{\alpha}{2\pi} Q_i^2
\left[
P_{q\to \gamma q}(z)
\left(
\log\left(\frac{M^2}{m_i^2}\right)-2 \log(1-z)-1
\right)
\right]_+
\nonumber\\
&&~~~~
+\bar f(z)
\nonumber
\eea
where $Q_i$ and $m_i$ are the electric charge fraction and the mass of the quark $i$;
the functions $f_i(z)~(i=q,\gamma)$~allow to change the subtraction scheme
(e.g. DIS or $\overline{MS}$) and are defined  as~\cite{ddh}
\bea
f_q(z) &=& 
2 \left(
\frac{\log(1-z)}{1-z}
\right)_+
-\frac32\frac{1}{(1-z)_+}
-(1+z) \log(1-z)
-\frac{1+z^2}{1-z}\log(z)\nonumber\\
&&~
+3 + 2 z -\left( \frac92 +\frac{\pi^2}{3} \right)
\nonumber\\
f_\gamma(z) &=&
\left(
(1-z)^2+z^2
\right)
\log\left(\frac{1-z}{z}  \right)
-8 z^2 + 8 z - 1
\eea
Since also the photon-induced processes contribute  to the
hadron-level cross section and develope a mass singularity when
the outgoing quark is collinear to the incoming photon,
also the Altarelli-Parisi splitting function $P_{\gamma\to q \bar q}$
contributes to the subtraction term for the quark densities.
The processes $\gamma q\to l^+ l^- q$,
because of a photon exchange in the $t$-channel in the
peripheral diagram,
develop a collinear singularity,
proportional to the splitting function $P_{q\to\gamma q}(z)$,
which can be reabsorbed in the photon density.
In the DIS scheme, the function $\bar f(z)$ can be fixed by any
combination of observables and will be therefore set to zero.

Given the presence in the hadron-level cross section of eq.~\myref{sigmahad} 
of the product of two parton densities, the subtraction procedure in a
factorized form could yield
terms of ${\cal O}(\alpha^2)$, which have been discarded  at \oa for 
consistency.

The generalization of the independence from the value of the
quark masses of the \oa cross section of eq.~\myref{sigmahad} to the
cross section including also QED higher-order corrections
has been discussed in detail in ref.~\cite{CMNV1} and
the main result is our master formula for the computation of
the hadron-level cross section and event simulation
\bea
\label{sigmahadsubinfty}
d\sigma_{had}&=&
\sum_{a,b=q,\bar q}\int_0^1 dx_1 dx_2~~
q_a(x_1,M^2) q_b(x_2,M^2)\times\\
&\Bigg\{&
{\tilde F}_{SV}~{\tilde\Pi}(Q^2,\varepsilon)~
\sum_{n=0}^\infty \frac{1}{n!}~
\left( \prod_{i=0}^n {\tilde F}_{H,i}\right)~
|{\tilde{\cal M}}_{n,LL}|^2~
d\Phi_n
\nonumber\\
&&~~+
\left[
d\sigma_\alpha^{ab,sub}
-
\left(
 \frac{\Delta q_a(x_1,M^2)}{q_a(x_1,M^2)}
+\frac{\Delta q_b(x_2,M^2)}{q_b(x_2,M^2)}
\right)
d\sigma_0^{ab}
\right]
\Bigg\}\nonumber\\
&&~~+
\, d\sigma_{had}^{q\gamma}
+
d\sigma_{had}^{\gamma\gamma}
\eea
where $d\sigma_{had}^{q\gamma}, d\sigma_{had}^{\gamma\gamma}$ can be
derived straightforwardly from eq.~(\ref{sigmahad}).

The factors with a tilde and also $d\sigma_\alpha^{ab, sub}$
represent quantities subtracted of the initial-state singularities
(cfr. Section 4 of ref. \cite{CMNV1} for the definitions and more details).

\section{Numerical results}
\label{results}
All the numerical results have been obtained using the following
values for the input parameters
\begin{center}
\begin{tabular}{lll}
$\alpha=1/137.03599911$ & 
$G_{\mu} = 1.16637~10^{-5}$ GeV$^{-2}$ &
$\mz=91.1876$~GeV \\
$\mw = 80.398$~GeV & 
$\gw = 2.4952$~GeV &
$\mh = 115$~GeV\\
$m_e=510.99892$~KeV &
$m_{\mu}=105.658369$~MeV &
$m_{\tau}=1.77699$~GeV \\
$m_u = 66$~MeV &
$m_c = 1.2$~GeV &
$m_t = 170.9$~GeV \\
$m_d = 66$~MeV &
$m_s = 150$~MeV &
$m_b = 4.3$~MeV \\
$V_{ud}=0.975$ &
$V_{us}=0.222$ &
$V_{ub}=0$ \\
$V_{cd}=0.222$ &
$V_{cs}=0.975$ &
$V_{cb}=0$ \\
$V_{td}=0$ &
$V_{ts}=0$ &
$V_{tb}=1$ \\
\end{tabular}
\end{center}
\begin{figure}
\begin{center}
\includegraphics[height=45mm,angle=0]{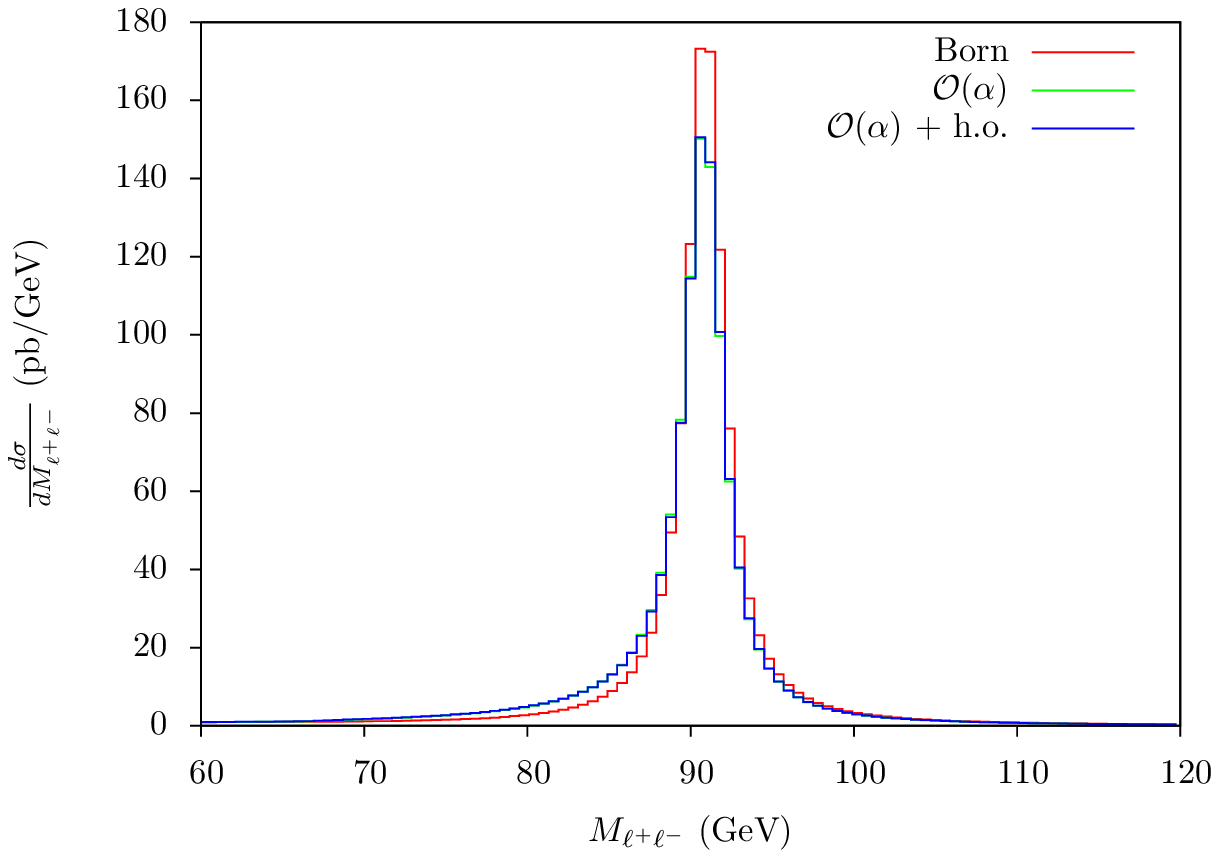}~
\includegraphics[height=45mm,angle=0]{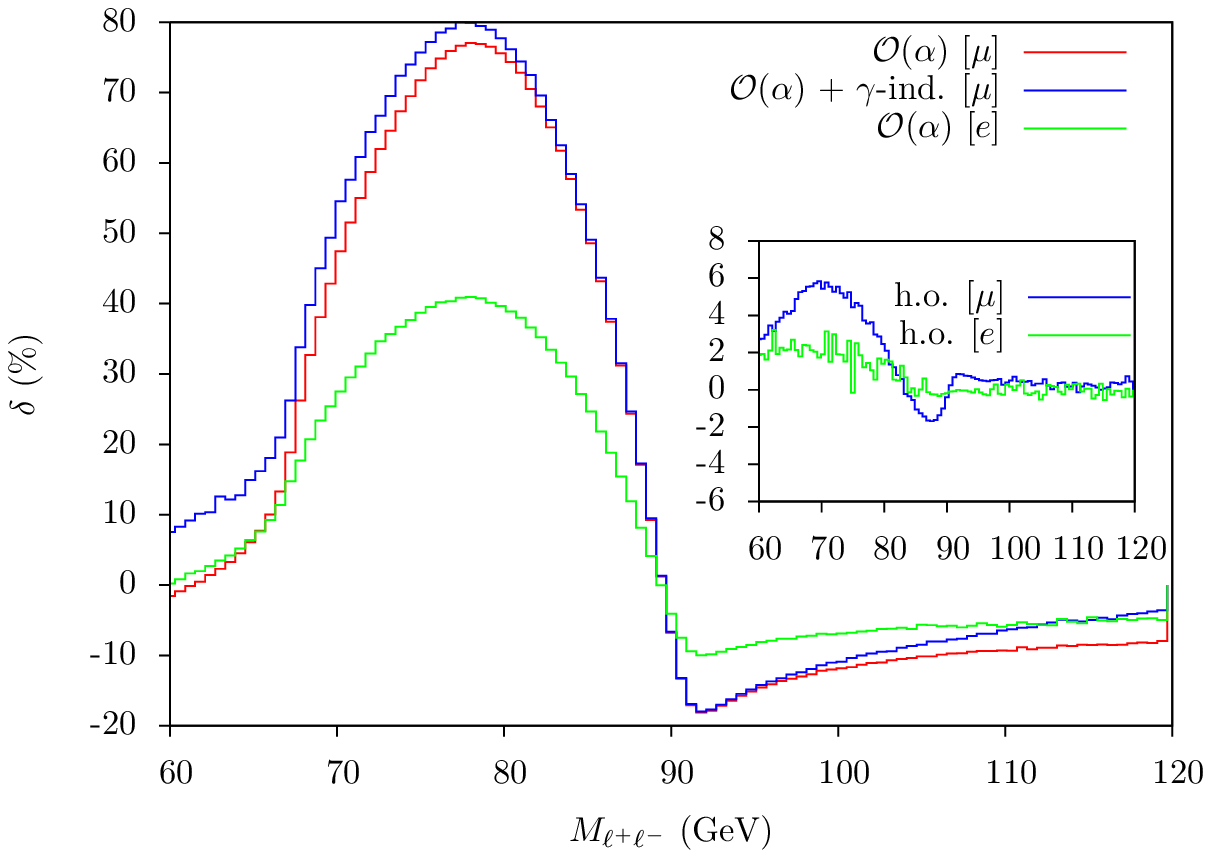}
\end{center}
\caption{
Invariant mass distribution around the $Z$ peak (left)
and
relative effect of different contributions (right), for bare muons and 
recombined electrons.
}
\label{minvdistrpeak}
\end{figure}
\begin{figure}
\begin{center}
\includegraphics[height=45mm,angle=0]{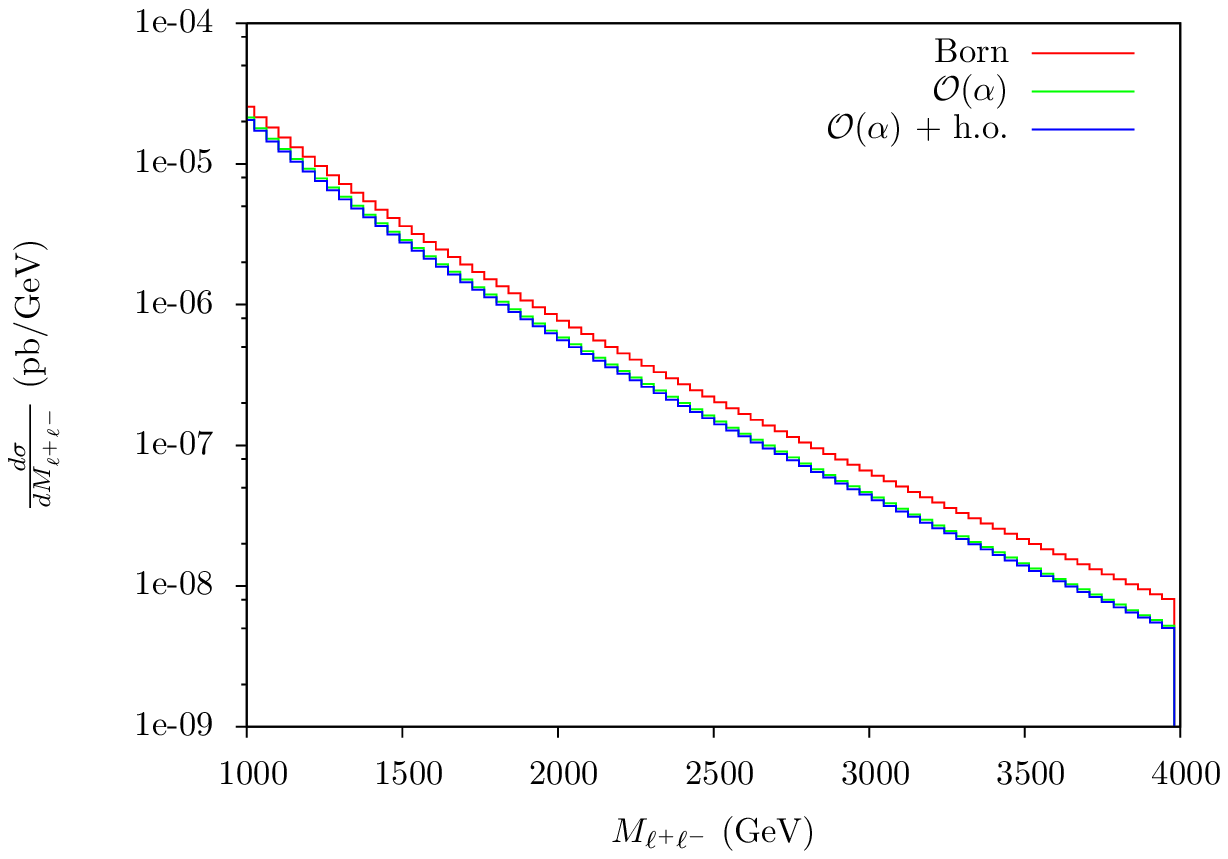}~
\includegraphics[height=45mm,angle=0]{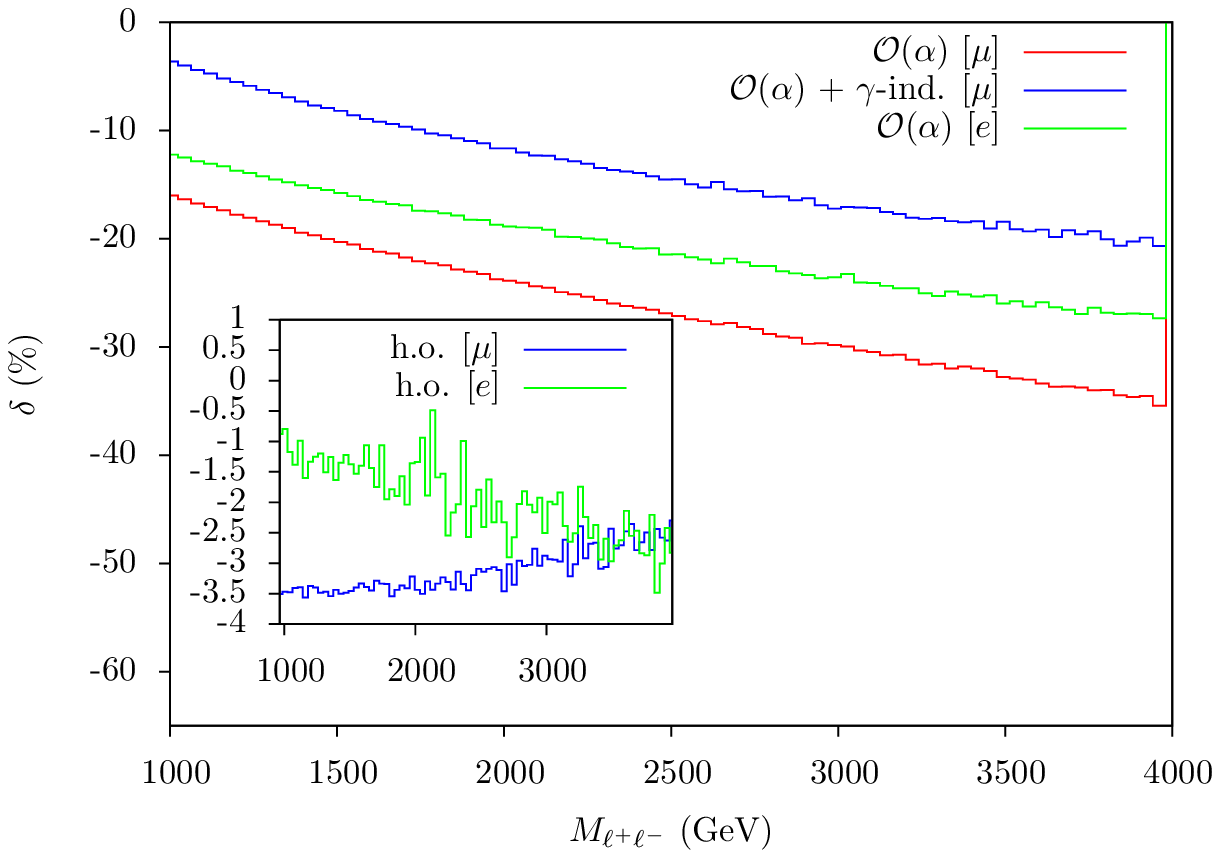}
\end{center}
\caption{
High tail of the invariant mass distribution (left)
and 
relative effect of different contributions (right), for bare muons and 
recombined electrons.
}
\label{minvdistrtail}
\end{figure}
and have been computed
in the EW scheme defined in Section~\ref{IBA}. 
The set of parton density functions used to compute all the hadron-level
cross sections is 
$\tt MRST2004QED$~\cite{MRST2004QED}.~
In this set of {\it pdf} the QCD and the QED factorization scales are set
to be equal and,
as usually done in the literature~\cite{BKS,BBHSW}, 
we choose $M=\mz$.
The use of the {\it pdf} set $\tt MRST2004QED$
implies that our numerical results  correspond to the DIS
factorization scheme.
The computation of the hadron-level results requires the numerical evaluation
of the subtraction term defined in eq. (\ref{deltaq});
a grid of values in the variable $x$, which is then interpolated,
is obtained by means of the numerical library $\tt CUBA$ \cite{CUBA}.
All the hadron-level results refer to the LHC, at a nominal
c.m. energy $\sqrt{s}=14$ TeV.

The following cuts have been imposed to select the events
\be
p_{\perp,\ell}>25~\mbox{GeV},\qquad
|\eta_{\ell}|<2.5
\label{cuts}
\ee
where $p_{\perp,\ell}$ and $\eta_\ell$ are the transverse momentum and
the pseudo-rapidity of the charged leptons, respectively.

Our results are obtained for muon pair final states. However, 
we also show results for recombined electrons in the case of the invariant
mass distribution.
In fact, we assume perfect isolation of photons from the muon, which is
experimentally achievable with good accuracy:
the resulting correction is
therefore amplified by large muon mass
collinear logarithms, because the photon emission is not treated
inclusively in the region around the muon.
In the case of electrons, it is not  possible experimentally to separate
them from the photon track, when the latter lies within a
cone around the lepton smaller than the detector angular resolution.
We adopt the following recombination
procedure
\begin{itemize}
\item photons with a rapidity $|\eta_\gamma|>2.5$ are
never recombined to the electron;
\item if the photon rapidity is $|\eta_\gamma|<2.5$ and
$R_{e\gamma}=\sqrt{(\eta_e-\eta_\gamma)^2+\phi_{e\gamma}^2 }<0.1$
($\phi_{e\gamma}$ is the angle between the photon and the
electron in the transverse plane), then the photon is recombined with the
electron, i.e. the momenta of the two particles are added and
associated with the momentum of the electron;
\item the resulting momenta should satisfy the cuts of
eq.~\myref{cuts}.
\end{itemize}

\begin{table}
\begin{center}
\begin{tabular}{|c|l|}
\hline
1. & lowest order (Born), as in Section \ref{IBA}, only $q{\bar q}$ subprocess\\
2. & exact \oa EW  corrections to the $q{\bar q}$ subprocess \\
3. & exact \oa EW  corrections to the $q{\bar q}$ subprocess plus 
$\gamma\gamma$ and $q\gamma $ contributions \\
4. & exact \oa EW matched with higher-order QED corrections to the
$q{\bar q}$ subprocess\\
\hline
\end{tabular}
\caption{Different approximations for the calculation of  the neutral current 
Drell-Yan cross section.}
\label{tableapprox}
\end{center}
\end{table}

\begin{table}
\begin{center}
\begin{tabular}{|c|c|c|c|c|}
\hline
           & 1.      & 2.        & 3.      & 4. \\
\hline
$\sigma({\rm pb})$  & 739.1(2)  &  710.7(1) & 715.8(1) & 712.8(2) \\
\hline
\end{tabular}
\caption{Cross sections, in pb, using approximations 1., 2., 3. and 4. given in table 1
and according to the EW scheme defined in section 2.3.}
\label{tablecrosssections}
\end{center}
\end{table}
In order to study the different effects of the radiative corrections and partonic
subprocesses on the relevant observables, we will distinguish the approximations
described in table~\ref{tableapprox}.

In table \ref{tablecrosssections} we compare the values
of the total cross section, within the cuts specified above, 
calculated in the approximations 1., 2., 3. and 4. of table
\ref{tableapprox}.
The effect of the \oa corrections is negative, of about 4\% of the
lowest-order result, while photonic subprocesses and QED higher orders
enhance the \oa cross section of about 1\% and a few per mille, respectively.

\begin{figure}
\begin{center}
\includegraphics[height=45mm,angle=0]{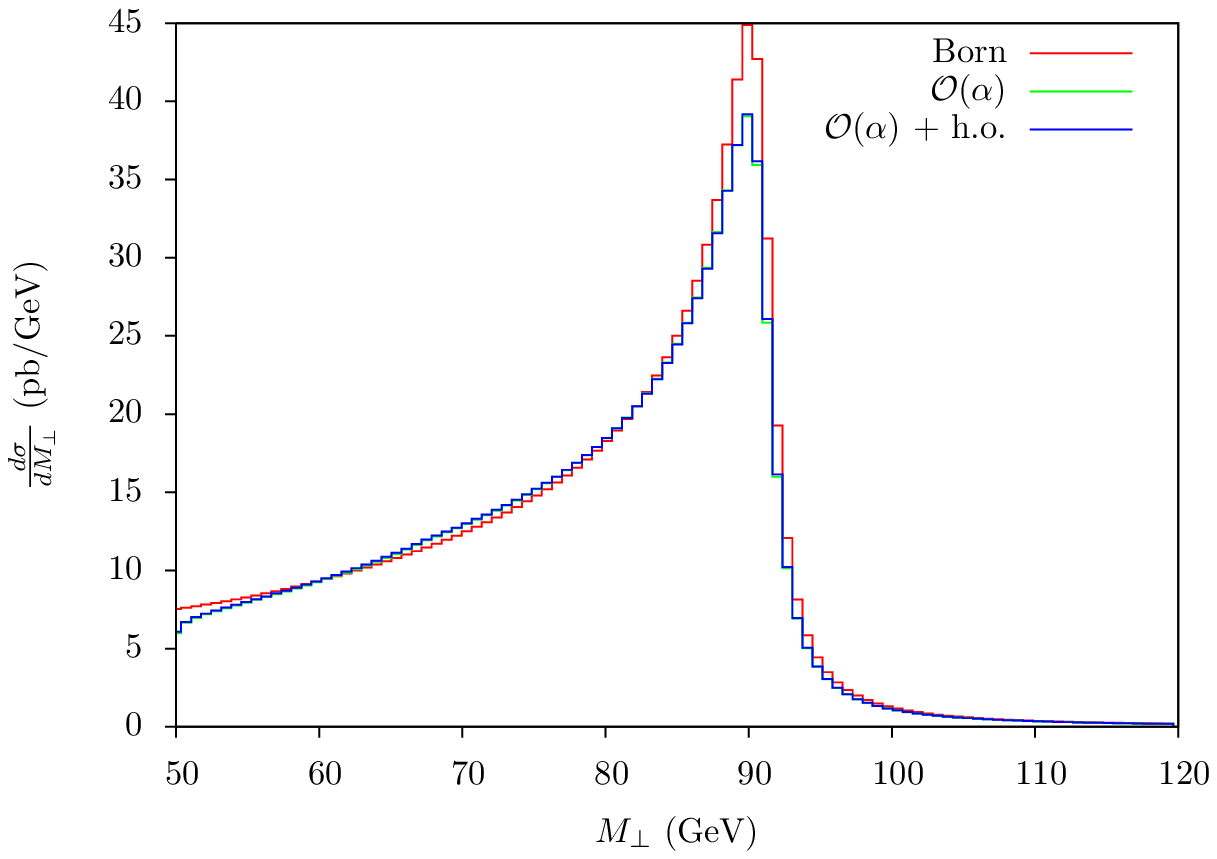}~
\includegraphics[height=45mm,angle=0]{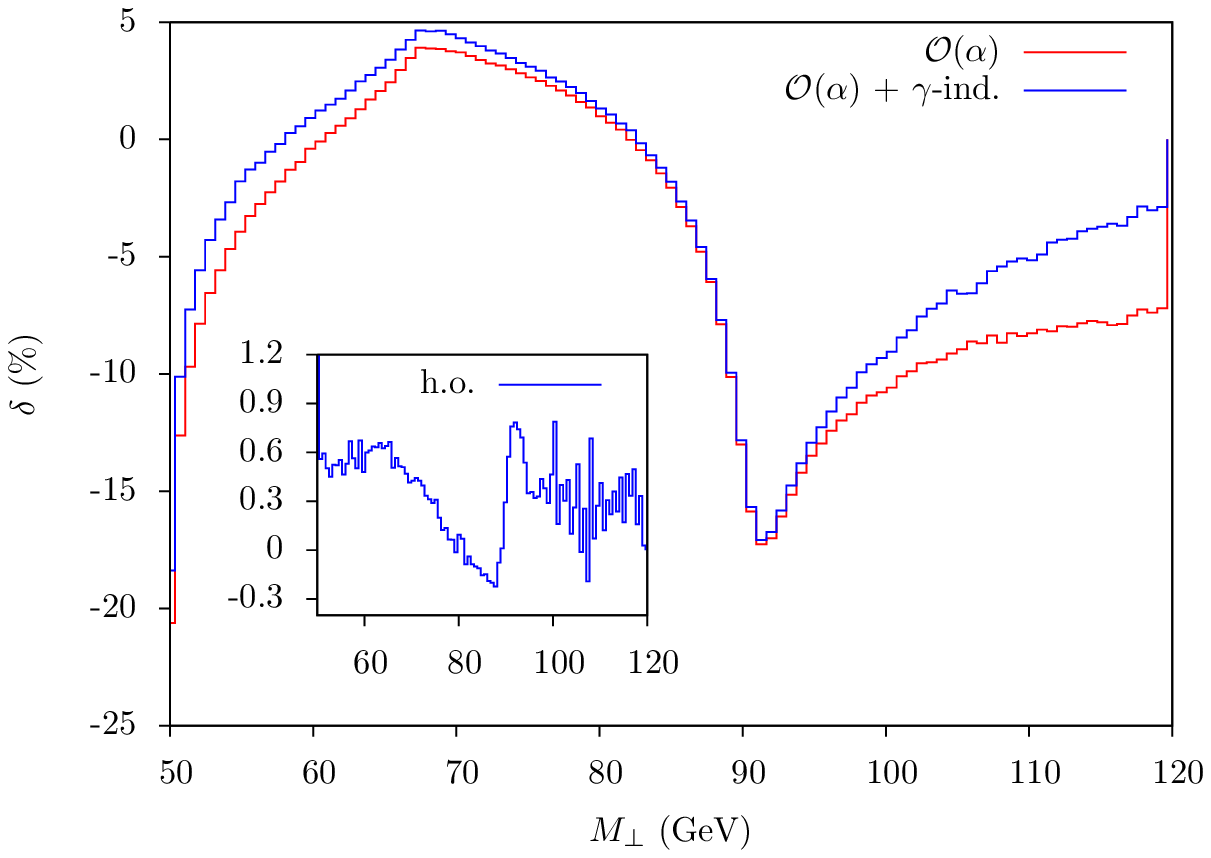}
\end{center}
\caption{
Transverse mass distribution around the $Z$ peak (left)
and
relative effect of different contributions (right).
}
\label{mtdistrpeak}
\end{figure}
\begin{figure}
\begin{center}
\includegraphics[height=45mm,angle=0]{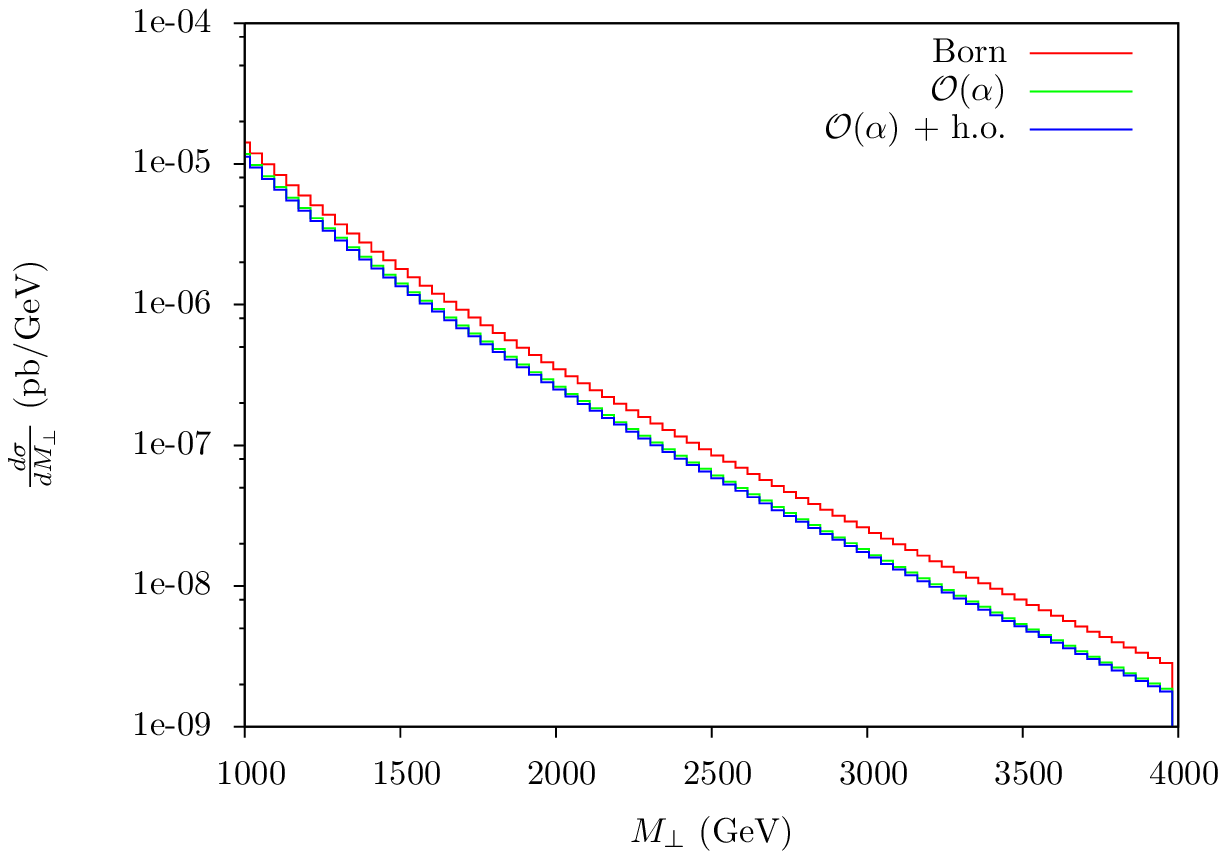}~
\includegraphics[height=45mm,angle=0]{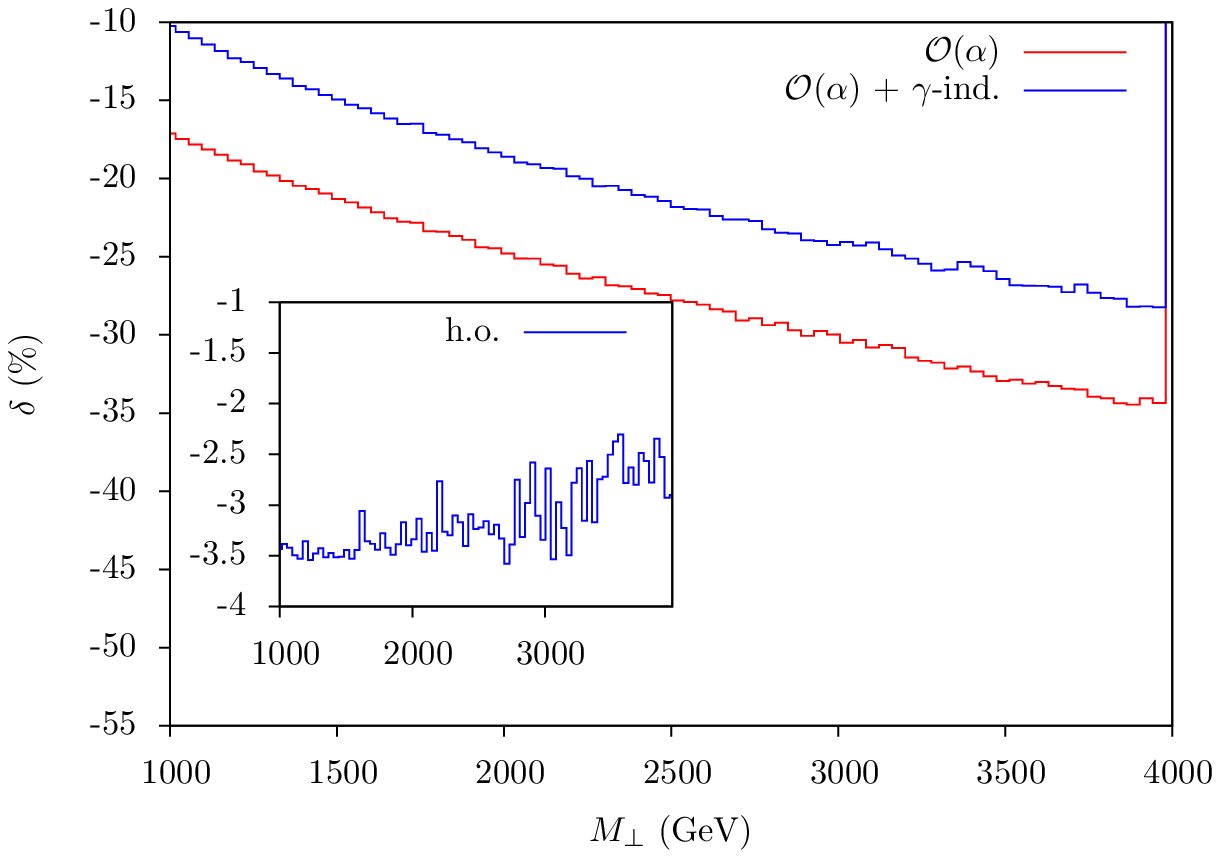}
\end{center}
\caption{
High tail of the transverse mass distribution (left)
and
relative effect of different contributions (right).
}
\label{mtdistrtail}
\end{figure}
\begin{figure}
\begin{center}
\includegraphics[height=45mm,angle=0]{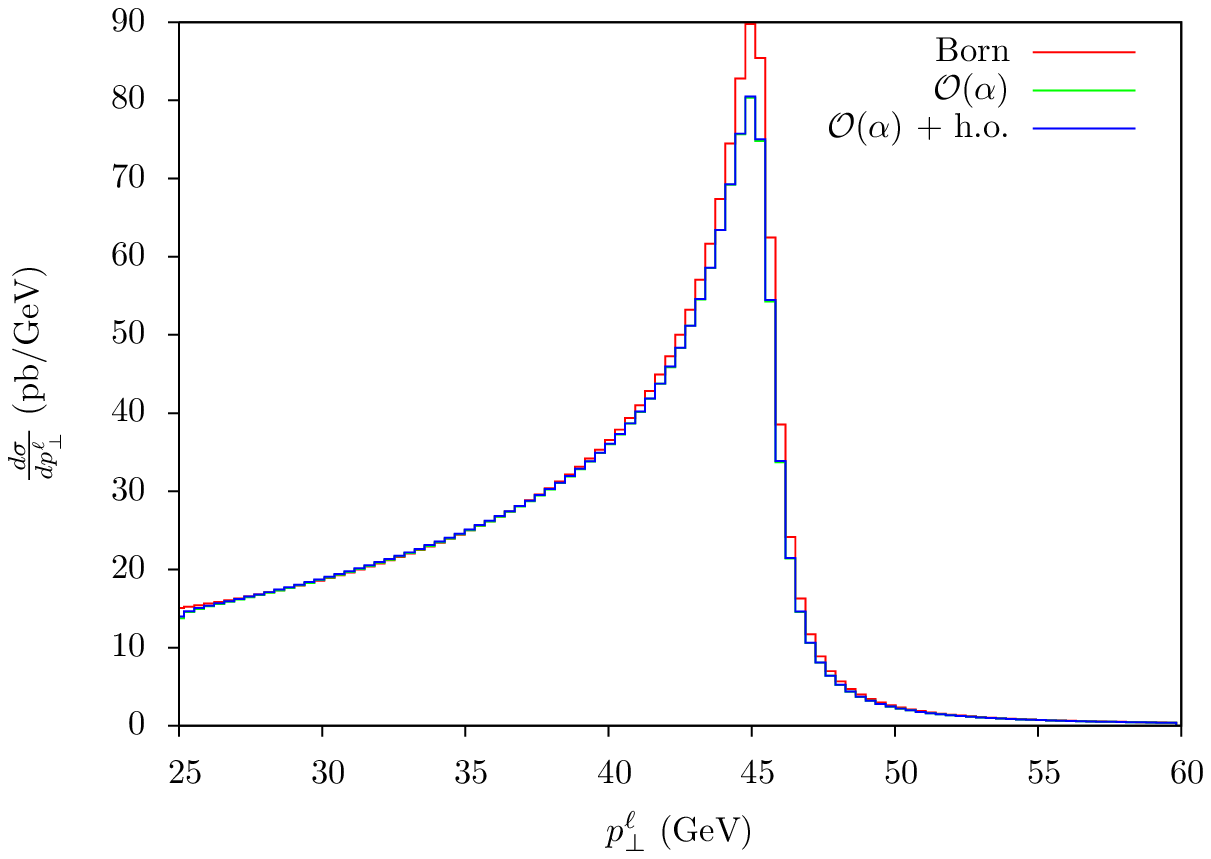}~
\includegraphics[height=45mm,angle=0]{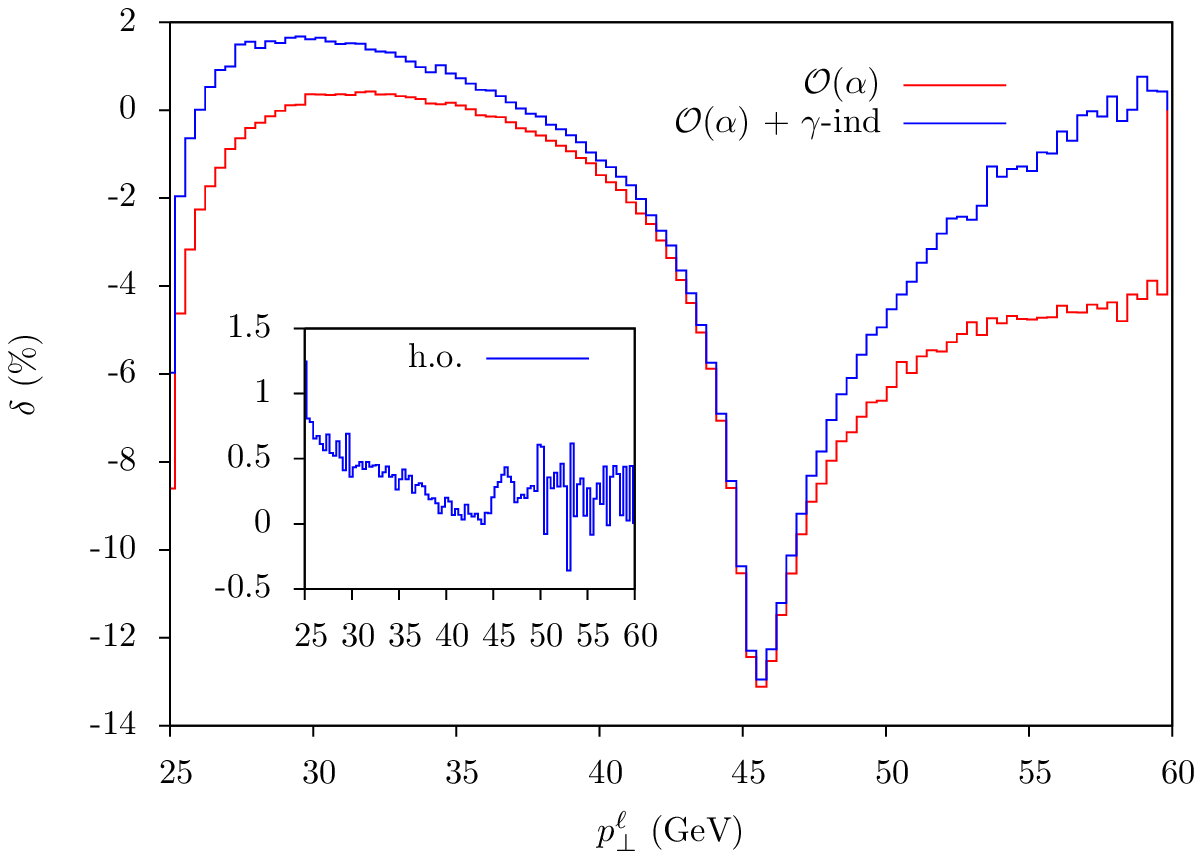}
\end{center}
\caption{
Lepton transverse momentum distribution around the $Z$ peak (left)
and
relative effect of different contributions (right).
}
\label{ptdistrpeak}
\end{figure}
\begin{figure}
\begin{center}
\includegraphics[height=45mm,angle=0]{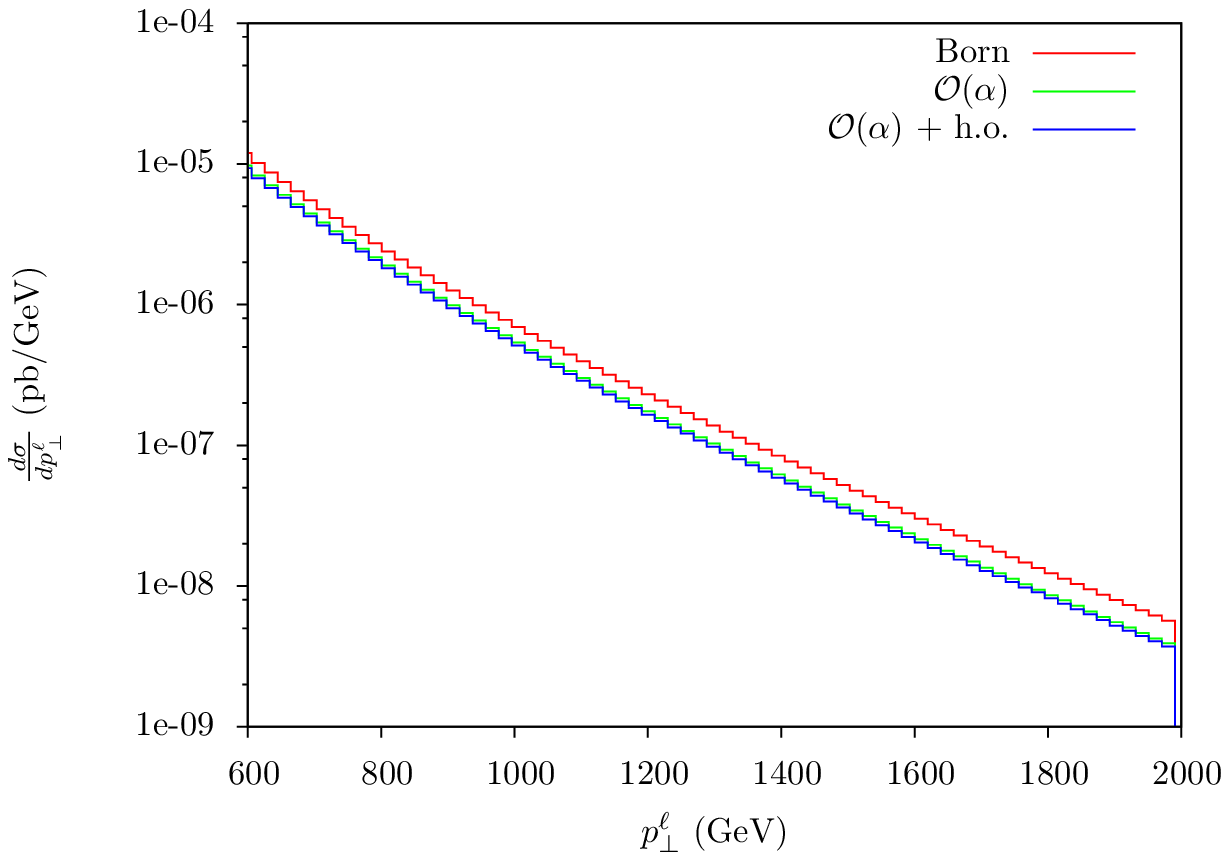}~
\includegraphics[height=45mm,angle=0]{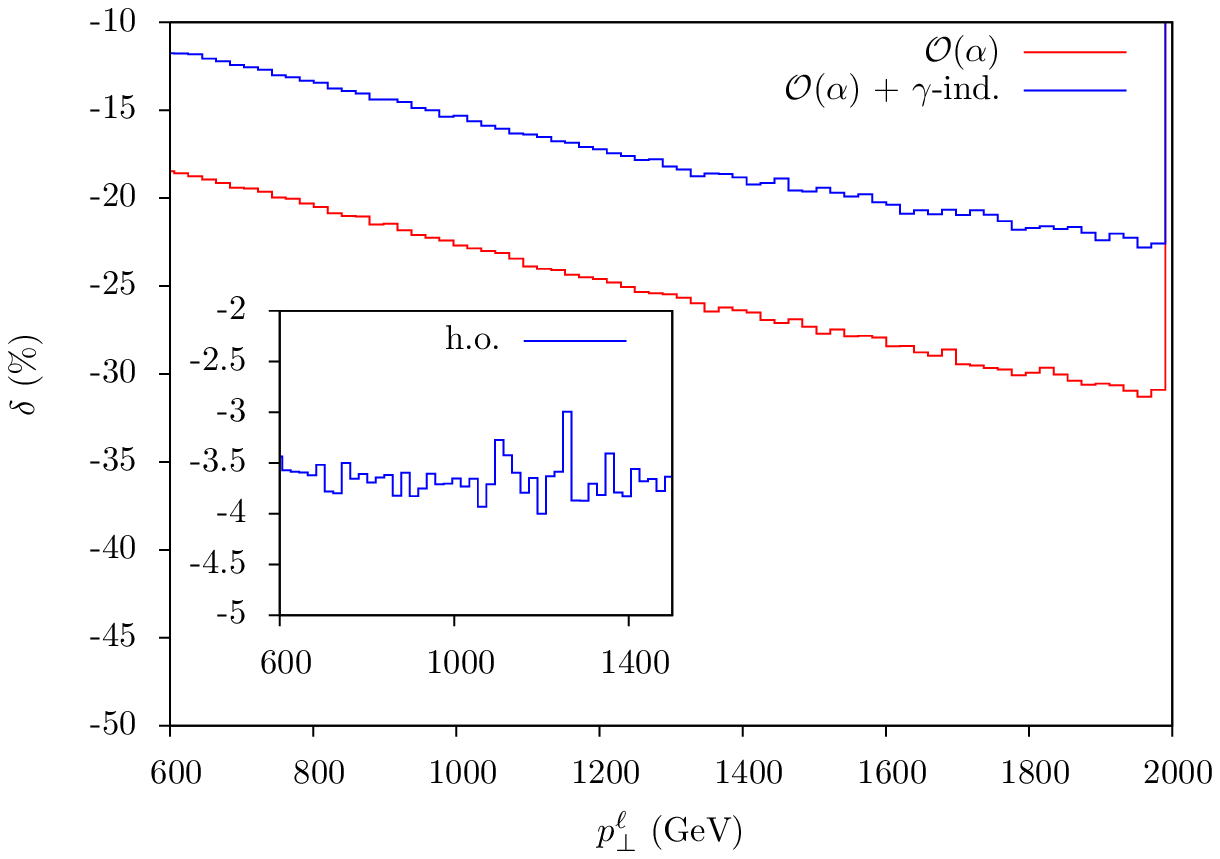}
\end{center}
\caption{
High tail of the lepton transverse momentum distribution (left)
and
relative effect of different contributions (right).
}
\label{ptdistrtail}
\end{figure}
\begin{figure}
\begin{center}
\includegraphics[height=45mm,angle=0]{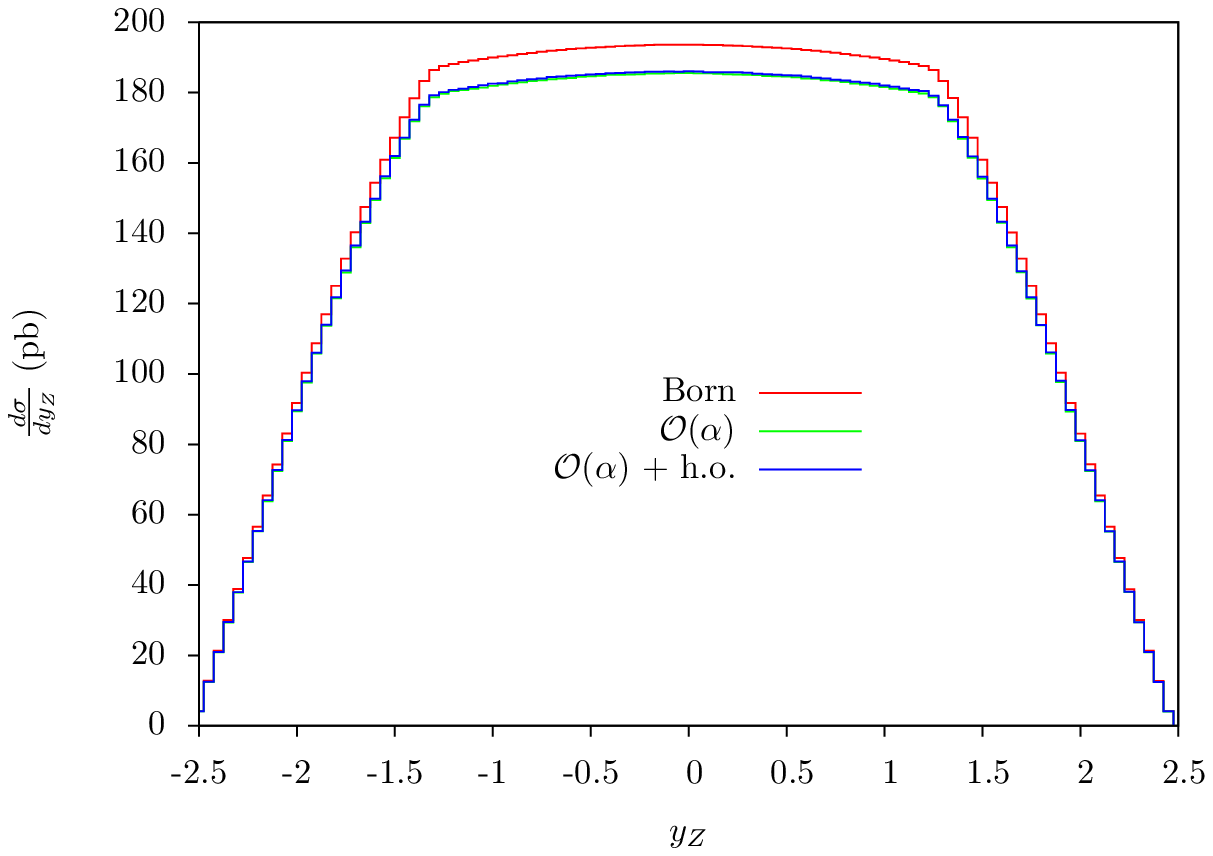}~
\includegraphics[height=45mm,angle=0]{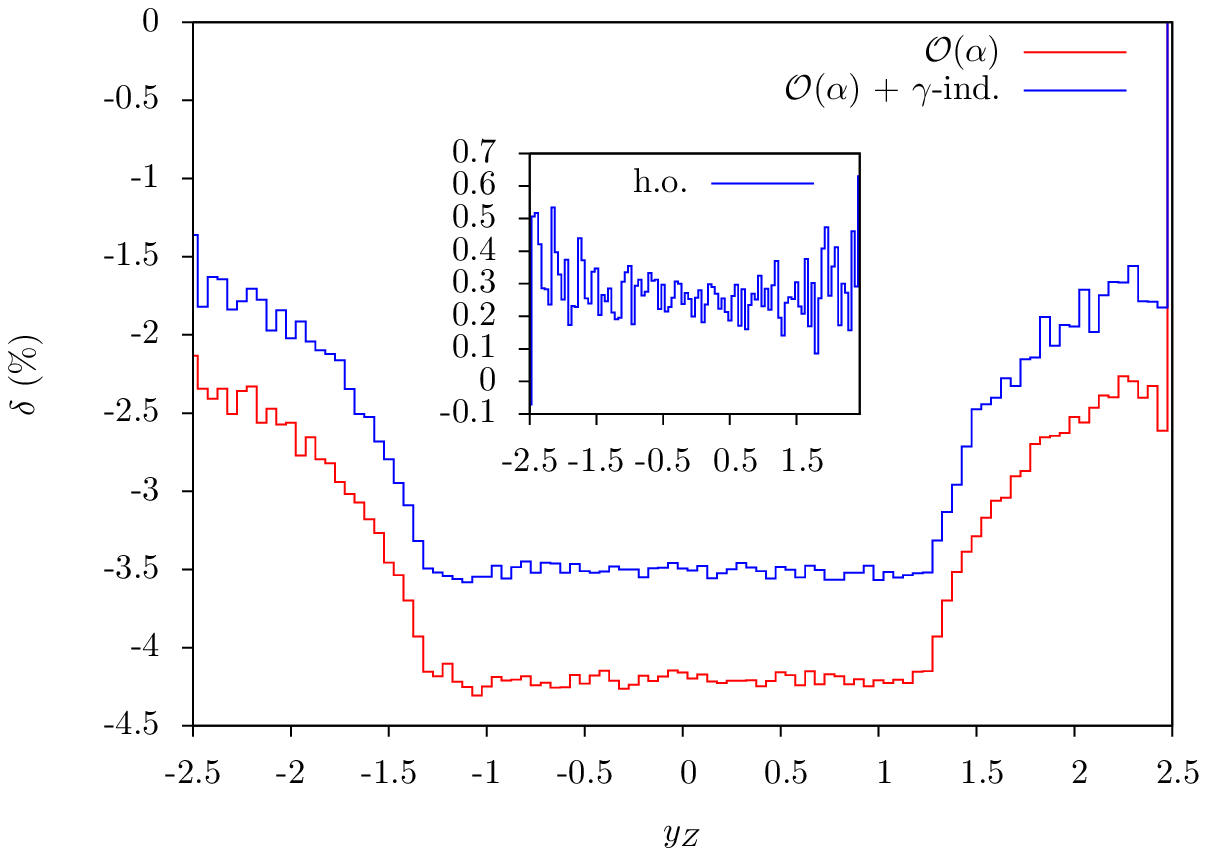}
\end{center}
\caption{
Rapidity distribution of the $Z$ boson (left)
and
relative effect of different contributions (right).
}
\label{yz}
\end{figure}

\begin{figure}
\begin{center}
\includegraphics[height=45mm,angle=0]{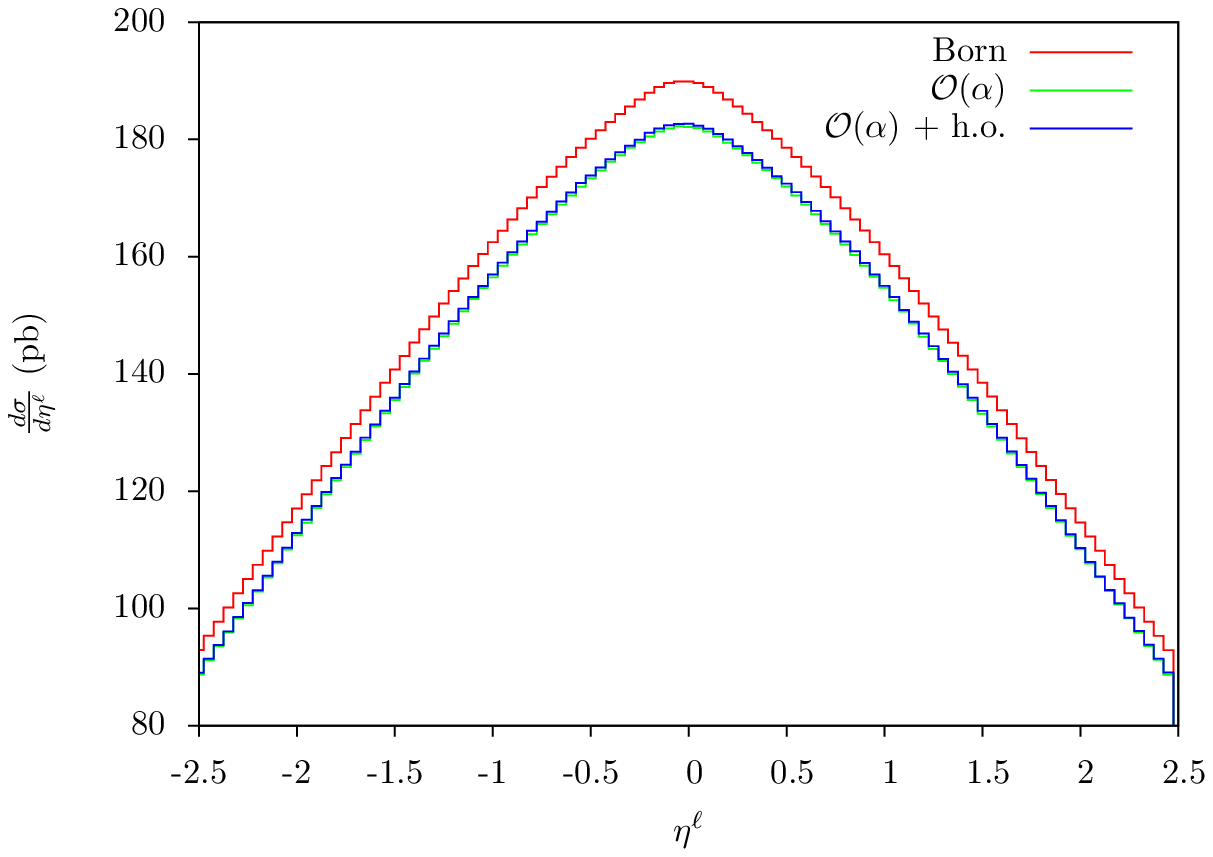}~
\includegraphics[height=45mm,angle=0]{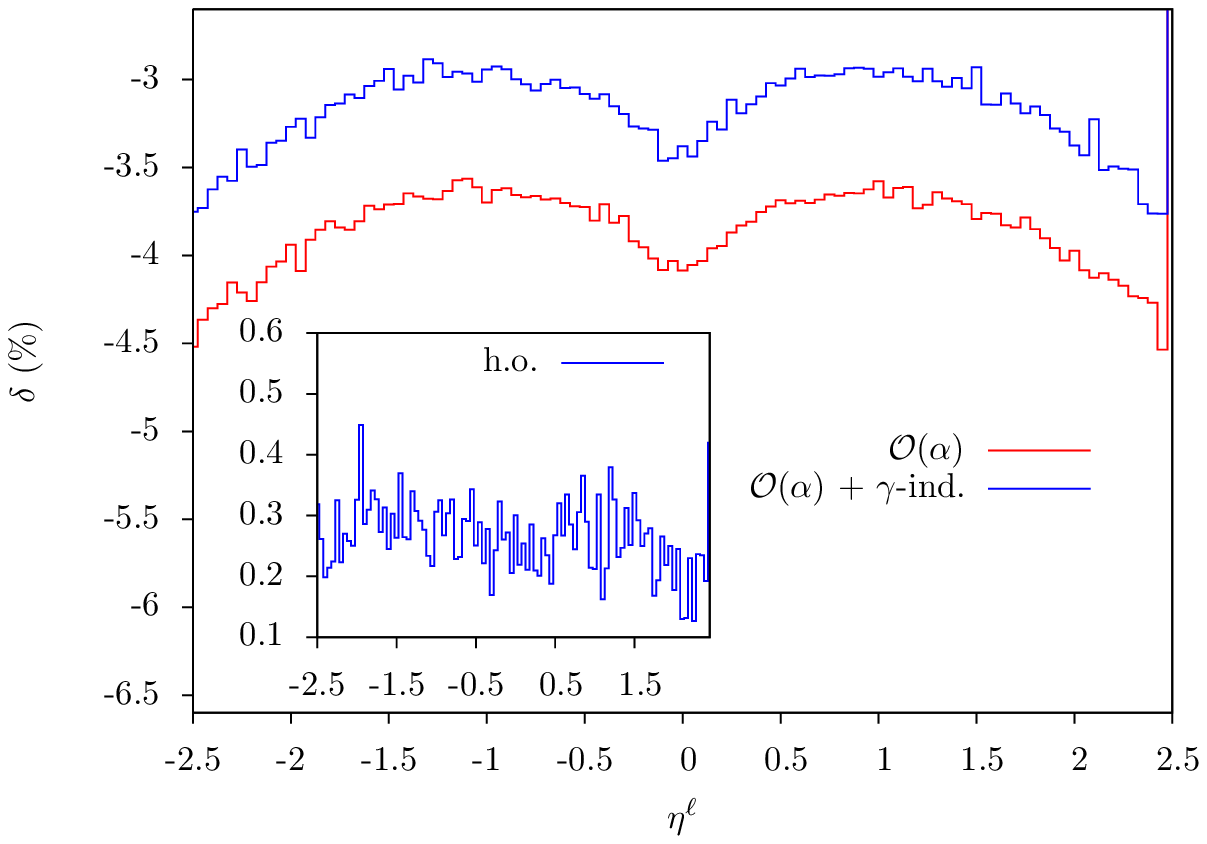}
\end{center}
\caption{
Pseudo-rapidity distribution of the final state lepton (left)
and
relative effect of different contributions (right).
}
\label{etal}
\end{figure}
\begin{figure}
\begin{center}
\includegraphics[height=45mm,angle=0]{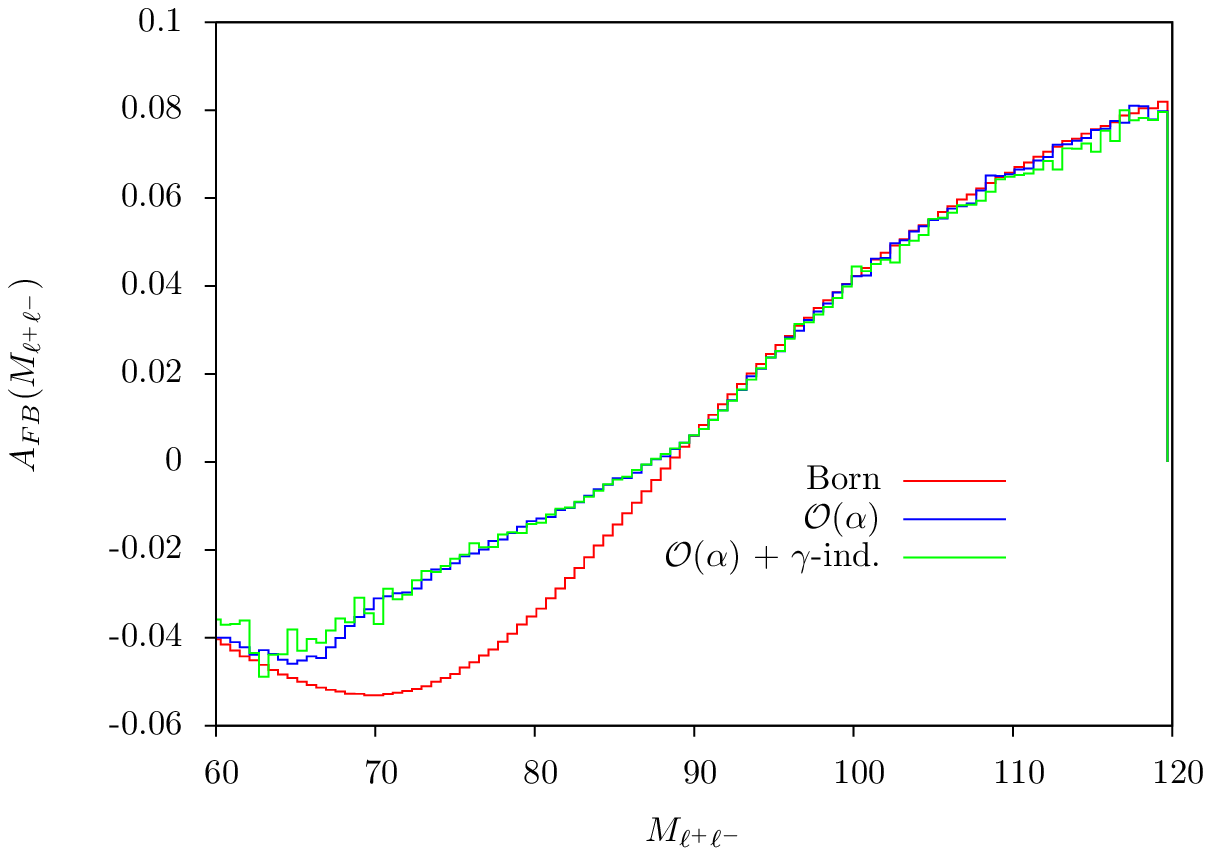}~
\end{center}
\caption{
Forward-Backward asymmetry as a function of the invariant mass of the
lepton pair according to the approximations 1., 2. and 3. of table 1.
}
\label{asym}
\end{figure}
In figure \ref{minvdistrpeak}
we show the invariant mass distribution of the final-state lepton pair, 
which is peaked at the $Z$-boson mass value and which,
taking as a reference the high-precision LEP measurement, 
can be used to calibrate the LHC detectors.
The radiative corrections significantly modify the shape of this
distribution, as can be seen from the right panel of figure \ref{minvdistrpeak}, 
showing the relative effect, expressed in units of the Born cross
section (approximation 1.), of the different contributions 
corresponding to approximations 2., 3. and 4. of table 1.
There is a well-known effect \cite{BKS,CMNT2} due to final-state photon 
radiation which increases, in the muon pair final state, 
by almost 80\% the lowest-order result
in the region below the $Z$ peak.
Around the peak, instead, the correction is negative of about $-15$\%.
The size of these corrections is significantly reduced when considering electron
final states, in agreement with the results known in the literature~\cite{BKS} and due to the photon recombination procedure,
which implies a partial cancelation of the mass logarithms, according
to the KLN theorem.
The impact of the photon-induced processes 
is of the
order of a few per cent and is particularly evident away from the resonance region.
The effect of multiple photon emission amounts to a few per
cent (see the inset of the figure) 
and can not be neglected for an accurate detector calibration and precision 
$Z$-physics studies at hadron colliders, as already remarked in ref.~\cite{CMNT2}.
The $Z$-boson mass shift induced by QED higher order corrections in fits to the
invariant mass distribution has
been already quantified in ref.~\cite{CMNT2} and found to be about 10\%
and of opposite sign of that caused by one-photon emission.

In figure \ref{minvdistrtail} we  show the
invariant mass distribution in the high tail region, 
where the Drell-Yan processes represent an important background to
the searches for new heavy gauge bosons.
The distribution receives large negative radiative corrections, of the order of
a few ten per cent, due to EW Sudakov logarithm, as previously emphasized in 
refs. \cite{BBHSW,Zyk}~\footnote{It has been recently noticed that weak boson emission 
diagrams, that contribute at the same order of \oa EW corrections, lead to a partial cancellation
of EW Sudakov corrections, when the weak boson decays into unobserved $\nu \bar{\nu}$ 
or jet pairs~\cite{BAUR}.}. 
With electron final states, the photon recombination procedure reduces the
negative effect of the \oa corrections. 
The multiple photon emission yields a few per cent effect in this invariant
mass region, while the photon-induced processes raise the cross section by approximately
+12\% of the Born approximation. 
The $\gamma\gamma\to l^+l^-$ subprocess contributes with about +3\% of
the $q\bar q$ Born cross section. Actually, its contribution is suppressed by the smallness of the photon density
and can be mainly observed at large invariant masses of the lepton pair.

In figure \ref{mtdistrpeak}  we consider  the distribution of
the $Z$ transverse mass, defined as
\be
M_\perp = \sqrt{2p_{\perp,l^+}~p_{\perp, l^-}~(1-\cos\phi_{l^+l^-})}
\ee
where $\phi_{l^+l^-}$ is the angle between the leptons in the transverse plane.
We remark the different size of the effect of the \oa corrections,
with respect to the invariant mass distribution, to this quantity, 
that can be useful, in association with the $W$
transverse mass distribution, to measure the $W$-boson mass. In the vicinity of 
the $Z$ resonance, the relative contribution of \oa EW corrections amounts to 
$\sim -17\%$ and is about a factor of two of the corresponding effect on the 
$W$ transverse mass distribution in the charged current channel \cite{CMNV1,DK}. This can 
be simply understood in terms of the dominance of final-state QED radiation within the 
full set of \oa EW corrections around the $Z$ resonance and, consequently, because of the presence of two radiating
leptons in the neutral current channel. Photon-induced processes reduce the relative size 
of \oa EW corrections and give a contribution of some per cent, especially above the
$Z$ peak, while multiple photon corrections contribute at some per mille level.

In figure \ref{mtdistrtail} we  show the high tail of the
transverse mass distribution, 
which also represents an important background observable to
the searches for new neutral gauge bosons.
The presence of the EW Sudakov logarithms reduces the cross section by
18 to 35\% for $1{\rm~TeV}\leq M_\perp \leq 4 $~TeV.
In contrast to the invariant mass case, the contribution of the
photon induced processes is smaller in this case, of the order of +7\%, 
while the effect of higher-order QED corrections is of the same size as that
observed in the high tail of the invariant mass distribution.

In figure \ref{ptdistrpeak}
the lepton transverse momentum distribution is shown around the $Z$
peak and, in figure \ref{ptdistrtail}, in the high momentum tail.
The effect of the radiative corrections, as well as of photon-induced processes,
 shows a pattern similar to the one observed for the transverse mass distribution and
 discussed above.

The rapidity distribution of the $Z$ boson is presented in figure
\ref{yz}. The \oa EW corrections are negative, of the order of -4\%,  
and almost constant in the interval $|y_Z|\leq 1$, 
and are smaller in size, at the 2\% level, for larger values of the rapidity. It is 
worth noting that  \oa EW contributions to such an
observables are of the same order of magnitude as NNLO QCD corrections 
\cite{ADMP} and, therefore, both the effects need to be taken into account 
in precision measurements of the $Z$ rapidity distribution.
We notice the overall positive correction due the photon-induced processes,
at the 1\% level.
The multiple photon emission has an almost negligible impact on this
observable, at the per mille level, as already discussed in ref. \cite{CMNT2}.

The numerical results for the pseudo-rapidity of the final state lepton
and the relative effect of the radiative corrections and photon-induced 
processes on this distribution are illustrated in figure \ref{etal}, showing a 
pattern quite similar to that observed for the $Z$ rapidity.

At hadron colliders it is possible to define a forward-backward
asymmetry $A_{FB}$ and to derive from it a measurement of the leptonic
effective weak mixing angle $\sin^2\theta_{eff}^l$. The forward-backward
asymmetry can be written as
\bea
A_{FB}(M_{l^+l^-}) &=& 
\frac{F(M_{l^+l^-})-B(M_{l^+l^-})}{F(M_{l^+l^-})+B(M_{l^+l^-})}\\
F(M_{l^+l^-})&=&\int_0^1d\cos\theta^*~\frac{d\sigma}{d\cos\theta^*}
~~~~~~~~
B(M_{l^+l^-})=\int_{-1}^0d\cos\theta^*~\frac{d\sigma}{d\cos\theta^*}
\nonumber
\eea
where
\bea
\cos\theta^*&=&f~\frac{2}{M(l^+l^-)
  \sqrt{M^2(l^+l^-)+p_\perp^2(l^+l^-)}}~
\left[  
p^+(l^-)p^-(l^+)-p^+(l^+)p^-(l^-)
\right]\\
p^\pm &=& \frac{1}{\sqrt{2}}(E\pm p_z),~~~~~~~~~~
f = 1~ ({\rm Tevatron}),~~~~~~
f = \frac{|p_z(l^+l^-)|}{p_z(l^+l^-)}~ ({\rm LHC})
\eea
and $M(l^+l^-)$ is the invariant mass of the final-state lepton pair, 
$p_\perp(l^+l^-)$ and $p_z(l^+l^-)$ are the total transverse momentum and
total longitudinal momentum of the $l^+ l^-$ pair, respectively. 
The asymmetry can be expressed in terms of $\sin^2\theta_{eff}^l$,
with good approximation, as $A_{FB} = b (a-\sin^2\theta_{eff}^l) $.
A detailed description of the effect of the \oa corrections on the
coefficients $a,b$ can be found  in ref.~\cite{BBHSW}, together with an
analysis of the relevant backgrounds.
In figure \ref{asym} we present the asymmetry distribution, evaluated
with the cuts of eq.~(\ref{cuts}), according to the approximations 1., 2. and 3. 
of table 1. It can be seen that \oa EW corrections are relevant and modify 
the shape of $A_{FB}$ below the $Z$ peak, whereas the photon-induced processes,  
as for the multiple photon  corrections \cite{CMNT2}, 
do not contribute significantly to this observable.

\section{Conclusions} 
\label{concllabel}
In this paper we have presented a precision electroweak calculation
of the neutral current Drell-Yan process.
The theoretical approach is based on the matching of exact \oa
EW corrections with QED Parton Shower, to account for the effect
of multiple photon emission.
The use of the proton {\it pdf} parametrization {\tt MRST2004QED}
allows for the presence of a photon density in the proton
and gives rise to additional partonic subprocesses which contribute to
the inclusive Drell-Yan signature.
The impact of the different contributions studied in the paper
turns out to be important, in association with higher-order 
QCD effects, in view of the precision measurements 
of EW parameters at the Fermilab Tevatron and the CERN LHC, as well 
as to validate the existing {\it pdf} parametrizations and to assess the SM
normalization in the search for new neutral heavy gauge bosons.

The calculation has been carried out using as inputs
$\alpha(0),\mw,\mz$ rather than a LEP-like choice
$\alpha(0),G_\mu,\mz$, in order to keep both gauge boson masses as
free parameters which can be fitted from the data,
since $W$- and $Z$-boson physics are intimately related at hadron colliders.

Possible perspectives of the present paper include
tuned comparisons with independent calculations, in the spirit of the
work done for single $W$ production during the Les Houches 2005~\cite{LH2005} and the
TeV4LHC~\cite{TeV4LHC} workshops,
a combination of EW and QCD corrections at the event generator level
and the addition of higher-order contributions such as 2-loop EW
Sudakov logarithms~\cite{SL}.

\section*{Acknowledgements}
We are grateful to Andreij Arbuzov and Dima Bardin for the 
numerical comparisons
at the preliminary stage of this work.
C.M.C.C. is supported by a ``Angelo Della Riccia'' fellowship
and thanks the CERN for the hospitality. He also thanks the Royal
Society and the British Council for partial support and is grateful
to the School of Astronomy and Physics of the Southampton University and
the NExT Institute where part of this work was done.
A.V. is supported by the European Community's Marie-Curie
Research Training Network under contract 
MRTN-CT-2006-035505 (HEPTOOLS).


%

\end{document}